%% file: main.tex
\def\inch#1{#1''}
\RequirePackage{fix-cm}
\newcommand{\orcid}[1]{\href{https://orcid.org/#1}{\includegraphics[width=8pt]{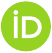}}}

\documentclass[twocolumn,epjc3]{svjour3}  
\smartqed  
\RequirePackage{graphicx}
\RequirePackage[numbers,sort&compress]{natbib}
\RequirePackage[colorlinks,citecolor=blue,urlcolor=blue,linkcolor=blue]{hyperref}
\RequirePackage{amsmath}
\RequirePackage{amssymb}
\RequirePackage{wasysym}
\RequirePackage{makecell}
\RequirePackage{siunitx}
\usepackage[switch]{lineno} 
\journalname{Eur. Phys. J. C}

\begin{document}

\title{Detection of scintillation light in noble gases with wavelength-shifting optical fibers}


\input{authors_epjc}





\date{Received: date / Accepted: date}

\maketitle

\begin{abstract}

Wavelength-shifting (WLS) techniques enable particle detectors based on noble gases, whose scintillation light is predominantly emitted in the vacuum-ultraviolet. We investigate WLS fibers coated with tetraphenyl butadiene (TPB) for scintillation light detection in gaseous xenon and argon at pressures up to 8.5~bar, motivated by future high-pressure xenon time-projection chambers of the NEXT program. Two detector configurations are studied: an elongated high-pressure vessel with four PTFE panels equipped with WLS fibers read by temperature-stabilized SiPMs, and a compact box-shaped detector operated at 1~bar Xe with WLS fibers read out by PMTs. Both operate with continuous gas purification. The detector response is characterized using cosmic muons and alpha particles from a $^{241}$Am source. With the SiPM setup, we measure a light collection efficiency (LCE) of ${1.18 \pm 0.01~\mathrm{(sta.)}~^{+0.07}_{-0.09}~\mathrm{(sys.)}~\%}$ for xenon and ${1.07 \pm 0.01~\mathrm{(sta.)}~^{+0.06}_{-0.08}~\mathrm{(sys.)}~\%}$ for argon. With PMT readout, we measure a LCE of ${0.45 \pm 0.01~\mathrm{(sta.)} \pm 0.05~\mathrm{(sys.)}~\%}$ in xenon, in agreement with the SiPM result once photon detection efficiency is accounted for. Average scintillation waveforms in xenon and argon are studied to assess the time structure of the emitted light. Cosmic-muon measurements yield a mean energy required to produce a scintillation photon $45\pm7~\mathrm{(sta.)}~^{+4}_{-5}~\mathrm{(sys.)}~\mathrm{eV}$ at 1.5~bar, in agreement with the literature. The results demonstrate that TPB-coated WLS fiber systems can reliably detect scintillation light in high-pressure gaseous noble detectors, with a LCE representing an upper limit for realistic large-scale TPCs, where additional photon losses from materials and fiber attenuation are expected.

\end{abstract}

\section{Introduction}

Noble elements such as xenon and argon are widely used in particle and astroparticle physics detectors thanks to their high scintillation yield, chemical inertness, and scalability to large target masses. They play a central role in experiments searching for rare events, including neutrino interactions~\cite{DUNE:2020lwj, Simon:2024cat}, dark matter scattering~\cite{XENON:2023cxc, LZ:2022lsv}, and neutrinoless double beta decay~\cite{EXO-200:2014ofj, NEXT:2023daz}. In particular, high-pressure gaseous xenon time projection chambers (HPGXe-TPCs) have emerged as a promising technology for next-generation neutrinoless double beta decay searches, combining excellent energy resolution with detailed topological event reconstruction~\cite{NEXT:2025ozn, NEXT:2020jmz}. The NEXT program has demonstrated the viability of this approach and is currently pursuing the development of progressively larger detectors towards the tonne scale~\cite{NEXT:2025yqw, NEXT:2020amj}.

A common challenge for detectors based on noble elements is efficient light collection, as scintillation photons are predominantly emitted in the vacuum ultraviolet (VUV) region, around 128 nm for argon and 175 nm for xenon, where most conventional photosensors are insensitive and common optical materials exhibit strong absorption. As a result, efficient photon detection typically requires the use of wavelength-shifting techniques to convert VUV light into the visible range.

A variety of solutions have been developed to address this challenge, including wavelength-shifting (WLS) coatings on large surfaces~\cite{DarkSide-20k:2017zyg, NEXT:2022cmg}, photosensors coupled to VUV-transparent windows~\cite{nEXO:2021ujk, Hawley-Herrera:2024evg}, and dedicated VUV light traps~\cite{Machado:2016jqe, Brizzolari:2021akq, Burgi:2024coo}. Among these approaches, WLS optical fibers offer several attractive features, including low material cost, reduced channel count through light collection over extended lengths, and the flexibility to instrument large or complex detector geometries. WLS fibers have already demonstrated excellent performance in liquid argon detectors such as ArgonCube~\cite{Anfimov:2020aev, DUNE:2024fjn}, as well as in experiments such as \textsc{Gerda} and \textsc{Legend}~\cite{Janicsko-Csathy:2010uif, GERDA:2022hxs, Burlac:2022bzq}, where they have been used to enhance background rejection.

The current NEXT-100 detector~\cite{NEXT:2025yqw} employs photomultiplier tubes (PMTs) for the detection of primary scintillation (S1) and electroluminescence (S2) signals. While this solution has proven successful, it presents significant challenges for future large-scale detectors. First, PMTs constitute one of the dominant sources of radioactive background. Second, they cannot operate directly in high-pressure xenon and must therefore be housed in a dedicated vacuum volume behind sapphire windows, separated from the active region by a thick copper pressure barrier. This configuration represents a major mechanical complication that becomes increasingly difficult to scale to larger detectors.

For this reason, future NEXT detectors are expected to rely exclusively on silicon photomultipliers (SiPMs), which are intrinsically radiopure and can operate directly in high-pressure xenon. The baseline design of NEXT-HD foresees SiPM tracking planes located at both ends of the TPC. However, while such planes are well suited for event topology reconstruction, using them simultaneously for S1 and energy measurement poses important challenges. The large number of channels required in tonne-scale detectors results in substantial data rates, particularly for the fast S1 signal. In addition, the dark count rate of room-temperature SiPMs makes the detection of low-energy S1 signals difficult, especially those produced by $^{83m}$Kr calibration events, which are essential for detector calibration and monitoring~\cite{NEXT:2018sqd, NEXT:2025fpq}.

To address these limitations, the NEXT collaboration is investigating the use of TPB-coated wavelength-shifting fibers as an alternative light-collection system~\cite{Soleti:2023sac}. In this concept, WLS fibers cover the barrel of the detector and collect both S1 and S2 light. The fibers are bundled and read out by a comparatively small number of SiPMs located outside the active volume. Since only a few hundred channels are required in a tonne-scale detector, these photosensors can be operated at high sampling rates and cooled to temperatures close to the xenon condensation point, reducing their dark count rate by approximately two orders of magnitude. This approach could enable efficient detection of low-energy S1 signals while maintaining excellent energy-measurement performance.

\begin{figure}[ht!]
    \centering
    \includegraphics[width=1\linewidth]{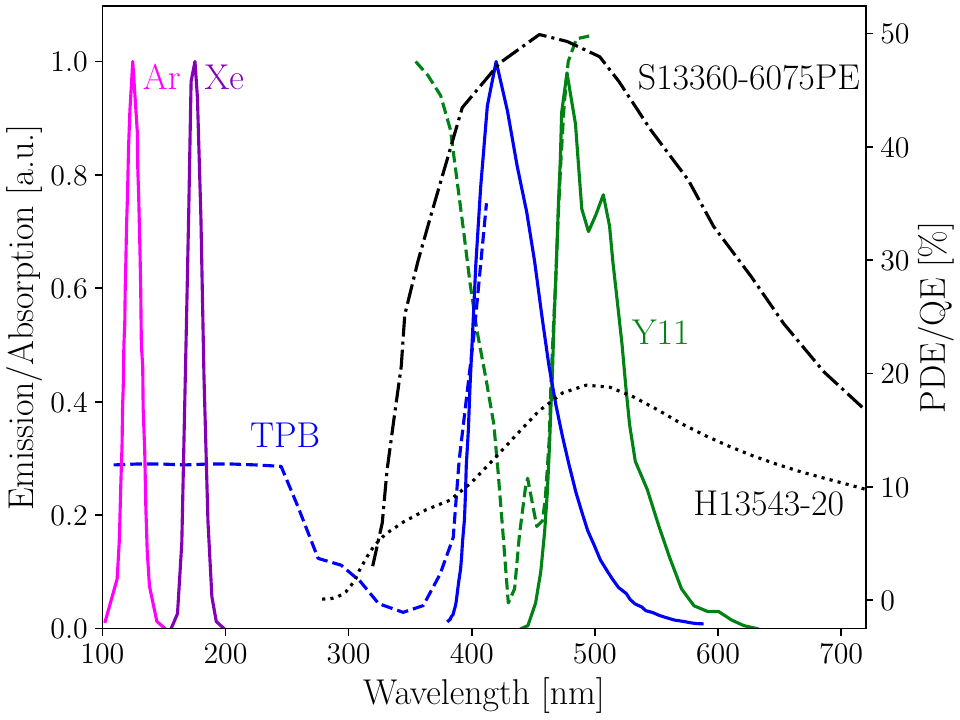}
    \caption{Emission (solid lines) and absorption (dashed lines) for xenon and argon (only VUV), TPB~\cite{GERDA:2022hxs}, and Y11 fibers~\cite{Kuraray}. The PDE of the SiPMs~\cite{Hamamatsu} (black dash-dotted line) and the QE of the red-enhanced PMT~\cite{Hamamatsu-red-PMT} (black dotted line) correspond to the right axis.}
    \label{fig:emission}
\end{figure}

The primary goal of this work is to assess the feasibility of such a light-collection system and to quantify the achievable light collection efficiency in gaseous noble elements. To this end, we developed two complementary detector prototypes equipped with TPB-coated WLS fibers. The first is an elongated chamber operated in xenon and argon gas at pressures between 1.5 and 8.5 bar and instrumented with SiPM readout. The second is a compact cubic detector operated in xenon at atmospheric pressure and read out with red-enhanced photomultiplier tubes. Figure~\ref{fig:emission} shows the emission spectra of xenon and argon, together with the TPB emission spectrum, the absorption spectrum of the Y11 fibers, and the photon detection efficiency of the SiPMs used in this study.

We determine the light collection efficiency using alpha-particle measurements in both setups and complement the xenon study with measurements based on cosmic muons. The elongated chamber is further used to characterize the detector response in gaseous argon over a wide pressure range. The results provide a quantitative assessment of the performance of TPB-coated WLS fibers in gaseous noble elements and contribute to the ongoing development of scalable light-collection solutions for future high-pressure noble-gas detectors, particularly tonne-scale HPGXe-TPCs.

The paper is organized as follows. In section~\ref{sec:setup}, both experimental setups are described, including the high-pressure vessel, the WLS fibers, the photosensor readout, and the gas and thermal control systems. Section~\ref{sec:methods} details the measurement procedure, data-taking conditions, and the calibration of the silicon photomultipliers. The experimental results are presented in section~\ref{sec:results}: alpha-induced scintillation measurements are used to determine the absolute light collection efficiency and to study the time structure of the scintillation signal, while complementary measurements with tagged cosmic-ray muons are employed to extract the scintillation light yield in xenon as a function of pressure. In section~\ref{sec:discussion}, the results are discussed in the context of electron-ion recombination effects in gaseous xenon and compared with existing measurements in the literature. Finally, section~\ref{sec:conclusions} summarizes the main conclusions and discusses the implications of this work for the design of future high-pressure gaseous detectors employing WLS fiber-based scintillation readout. 

\begin{figure*}[ht!]
    \centering
    \includegraphics[width=0.75\linewidth]{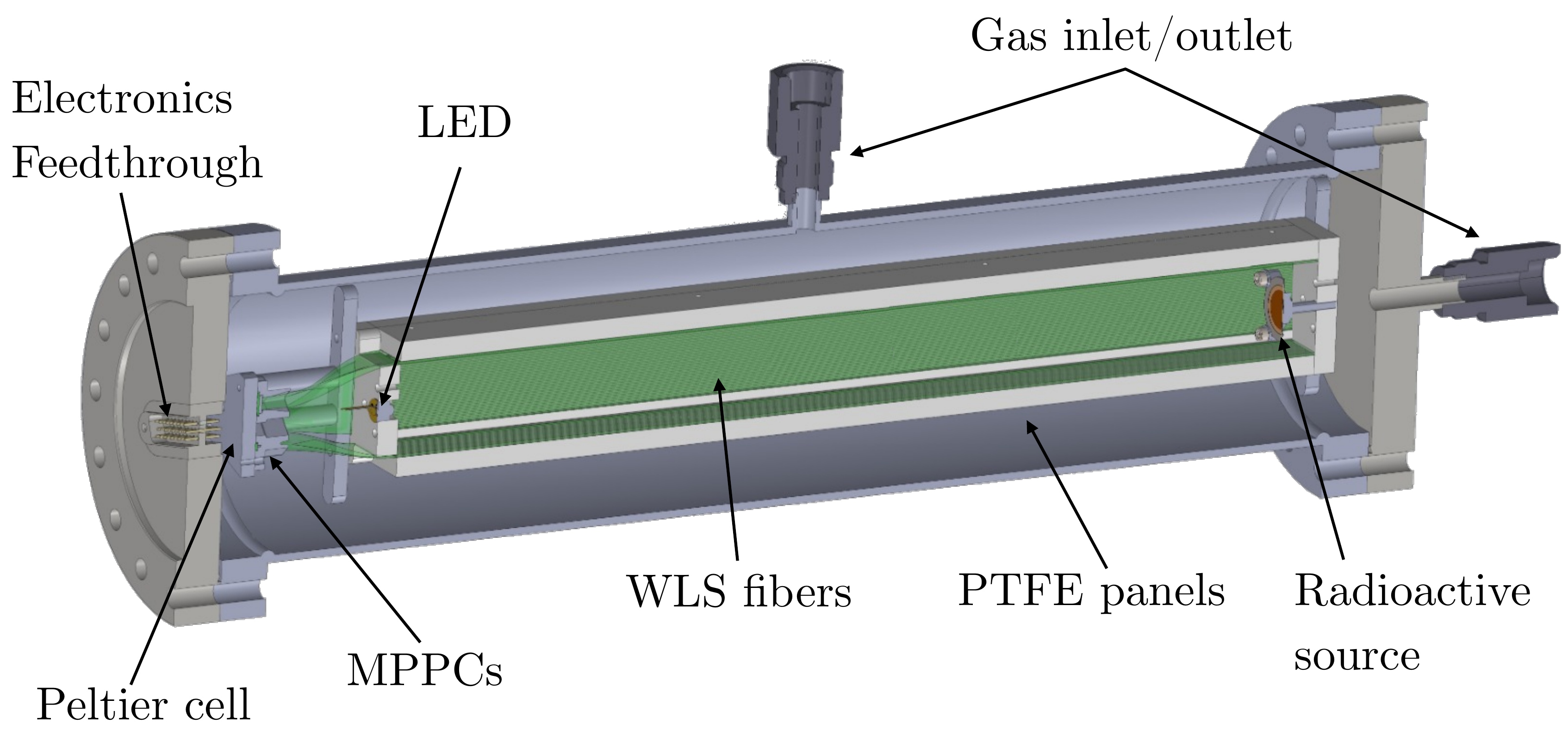}
    \caption{CAD cutaway view of the elongated experimental setup. Four PTFE panels instrumented with TPB-coated wavelength-shifting fibers and sealed at both ends with matching PTFE endcaps define the active volume inside a high-pressure vessel. The fiber bundles covering the panels are read out by Hamamatsu MPPCs mounted on a copper plate and connected to custom front-end electronics, with a Peltier-based cooling system located behind the plate for thermal stabilization. Gas inlet and outlet ports, as well as feedthroughs for electronics, an internal LED, and a radioactive source, are also shown. The two plastic scintillator muon taggers are not shown here.}
    \label{fig:setup}
\end{figure*}

\section{Experimental setups}\label{sec:setup}

Fig.~\ref{fig:setup} displays the setup with SiPM readout, which consists of an elongated horizontal high-pressure vessel instrumented with an internal light-collection system based on WLS optical fibers. The active zero-field volume is defined by four polytetrafluoroethylene (PTFE) panels, each with dimensions of $350 \times 45 \times 8~\mathrm{mm}^3$, sealed with matching PTFE endcaps and arranged to form a closed square geometry with an inner cross section of approximately $37 \times 37~\mathrm{mm}^2$ over a length of 350~mm. Each panel is covered with 30 Kuraray Y11 WLS multi-clad fibers~\cite{Kuraray} of 1~mm diameter and about 420~mm length, fixed onto the PTFE surface with three thin epoxy strips and coated with an approximately 3~\si{\micro\meter}-thick layer of tetra\-phenyl butadiene (TPB) to convert VUV scintillation light to visible wavelengths~\cite{Gehman:2011xm, NEXT:2022cmg}. One end of each fiber is aluminized, providing a reflectivity of approximately 75\%~\cite{Saraiva:2004cn} to enhance light collection. The opposite ends are grouped together in square bundles of 6$\times$6 mm$^2$ and coupled with optical glue to a Hamamatsu MPPC S13360-6075PE~\cite{Hamamatsu} silicon photomultiplier with an active area of the same size of the bundles, resulting in one photosensor readout per panel.

The four SiPMs are mounted on a common copper plate that provides both mechanical support and efficient thermal conduction. The four-panel assembly is housed inside a cylindrical stainless-steel vessel of 100~mm diameter and 500~mm length, equipped with DN100CF flanges at both ends and two \inch{$1/2$} VCR gas ports, one located at the center of the vessel and a second mounted on a flange. Two EJ200 plastic scintillator~\cite{eljen} panels are used to tag cosmic rays, with active areas of $15 \times 15~\mathrm{cm}^2$ and $10 \times 10~\mathrm{cm}^2$, respectively, and are positioned below and above the pressure vessel to provide a coincidence trigger for through-going muons. 

The detector is connected to a dedicated gas system that continuously purifies the lab-quality gas (purity-grade 5) using a hot getter from Gatekeeper\textsuperscript{\tiny\textregistered} model PS4-MT15~\cite{SAES}. The SiPMs are connected to a custom front-end electronics board that provides two different voltage gains ($1\times$ and approximately $300\times$) and a common bias voltage to all four channels, which is then interfaced to a CAEN DT2740 digitizer~\cite{CAEN} for data acquisition, operating at 16~bits with a sampling time of 8~ns.

Thermal stabilization of the photosensors is achieved by a cooling system placed behind the copper plate, consisting of a Peltier cell coupled to a radiator and a small fan. These elements are interfaced to a PID-based temperature control system (RS PRO 154 ESM 4420) and are used to actively cool and stabilize the SiPMs during operation. The operating temperature range varies from 14.0°C to 6.0°C, depending on the system's heat dissipation in the gas at different pressures and the ambient laboratory conditions. Within this temperature range, the SiPM dark count rate goes from 2.5 MHz to 1 MHz, approximately. A holder for a $^{241}$Am radioactive source is installed on the inner surface of the endcap located on the opposite side of the photosensors. The source is formed by electroplating, such that the maximal energy of the $^{241}$Am alpha is close to 5.49 MeV and has an activity of 2.9 kBq.

\begin{figure}
    \centering
    \includegraphics[width=0.9\columnwidth]{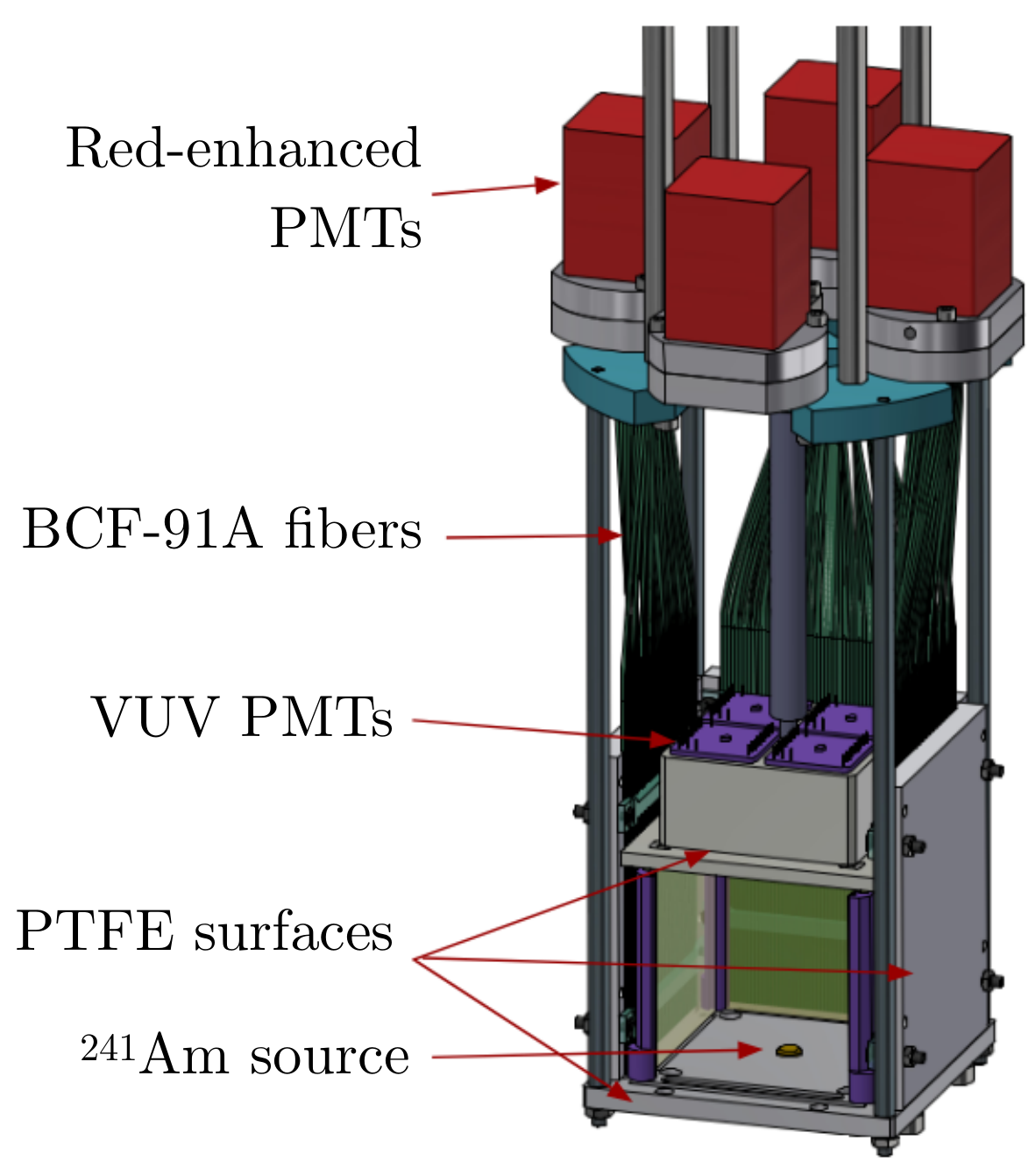}
    \caption{Cutaway view of the box-shaped experimental setup. An approximately cubic volume is defined by four PTFE panels instrumented with 64 TPB-coated wavelength-shifting fibers each, a PTFE floor holding an $^{241}$Am source at its center and a PTFE ceiling structure holding up to 4 VUV PMTs. The fibers are extracted to the top, where they are optically coupled to 4 red-enhanced PMTs.}
    \label{fig:frogxe-setup}
\end{figure}

Fig.~\ref{fig:frogxe-setup} shows the setup with PMT readout. The apparatus consists of a xenon-filled volume ($74 \times 74 \times 84~\mathrm{mm}^3$) defined by PTFE panels. The vertical walls are instrumented with 64 round multi-clad WLS optical fibers of 1~mm diameter each (Luxium, formerly Saint-Gobain, BCF-91A~\cite{Luxium}), aluminized at one end. The fibers are held in place by two acrylic supports and cover approximately 87\% of the PTFE surface. Each fiber is coated with a $\sim$3~\si{\micro\meter} layer of TPB, deposited by physical vapor deposition. The bottom panel hosts a 7~mm diameter $^{241}$Am source at its center, while the top panel consists of a PTFE structure capable of accommodating up to four VUV-sensitive photomultiplier tubes (Hamamatsu R8520-406 \cite{Hamamatsu-VUV-PMT}). The source (with a nominal activity of 140 kBq) is foil-type, with a 1-\si{\micro\meter}-thick, 5-mm-diameter active matrix of americium oxide in gold, covered by a 2-\si{\micro\meter}-thick protective layer of palladium, resulting in a peak alpha energy of 4.6 MeV (determined by a dedicated measurement using a solid-state spectroscopic alpha detector in vacuum). For the measurements reported here, only one PMT was installed; the remaining surface was covered with Vikuiti\textsuperscript{\texttrademark} Enhanced Specular Reflector. The WLS fibers are routed to the upper section of the apparatus, where each bundle is optically coupled to a red-enhanced PMT (Hamamatsu H13543-20 \cite{Hamamatsu-red-PMT}). The entire assembly is supported from a flange and housed within a stainless-steel vessel filled with xenon at a pressure of 1 bar.

\section{Measurement procedure}\label{sec:methods}
In the SiPM-based setup, each data-taking run begins with switching on the hot getter of the gas system approximately one hour before the start of the acquisition, allowing sufficient time for gas purification and stabilization. 

Data were acquired at several operating pressures between 1.5 and 8.5~bar, with a pressure accuracy of  $\sim0.1$~bar level. For all runs the \(\mathrm{^{241}Am}\) source was positioned on one side of the active volume. In the case of the xenon runs, a second data taking was performed using coincidence triggers from plastic scintillator panels placed above and below the stainless-steel vessel, to select through-going cosmic muons. At the start of each data taking, the temperature controller, cooling fan, and Peltier cell are switched on, and the SiPM bias voltage is adjusted accordingly to account for the breakdown voltage variation with temperature. Since the minimum achievable temperature depends on the operating pressure (higher pressures resulting in increased heat dissipation), the photosensor temperature and corresponding bias voltage are optimized and set independently at each pressure point. Thus, the SiPMs calibration is performed at each temperature and pressure, with the procedure described below.


The waveforms are reconstructed in a consistent manner throughout the entire analysis. The reconstruction procedure begins with the determination of the baseline, followed by baseline subtraction. For each run, the distribution of the pre-trigger samples is constructed, and the core of this distribution is fitted with a Gaussian function. This approach is adopted because of the high pileup probability in the amplified channel coming from dark current photoelectrons, which create a long tail in the distribution. 

For the PMT-based setup, we follow an equivalent procedure. The acquired data corresponds to alpha interactions in the gas at a fixed pressure of 1 bar with no electric field. The data presented here were also taken after continuous circulation through a hot getter (SAES~PS4-MT3-R-2 \cite{SAES}) over a weekend. The light yield was monitored throughout this period and was observed to reach a stable plateau after a few hours. 

The data processing was equivalent to that of the SiPM setup. The waveforms recorded by the four red-enhanced PMTs were baseline-subtracted and integrated over 2~\si{\micro\second} to obtain the total collected charge. This charge was converted to photoelectrons using the gain determined from a single-photoelectron calibration~\cite{Dossi:1998zn}.

\subsection{SiPM calibration}\label{sec:mppc}
The calibration of the SiPMs is performed for each data-taking run and for each operating temperature. The high-gain channel provided by the front-end electronics board is used to trigger on the intrinsic dark pulses of the SiPMs, enabling the acquisition of single-photoelectron signals.

\begin{figure}[ht!]
    \centering
    \includegraphics[width=1\linewidth]{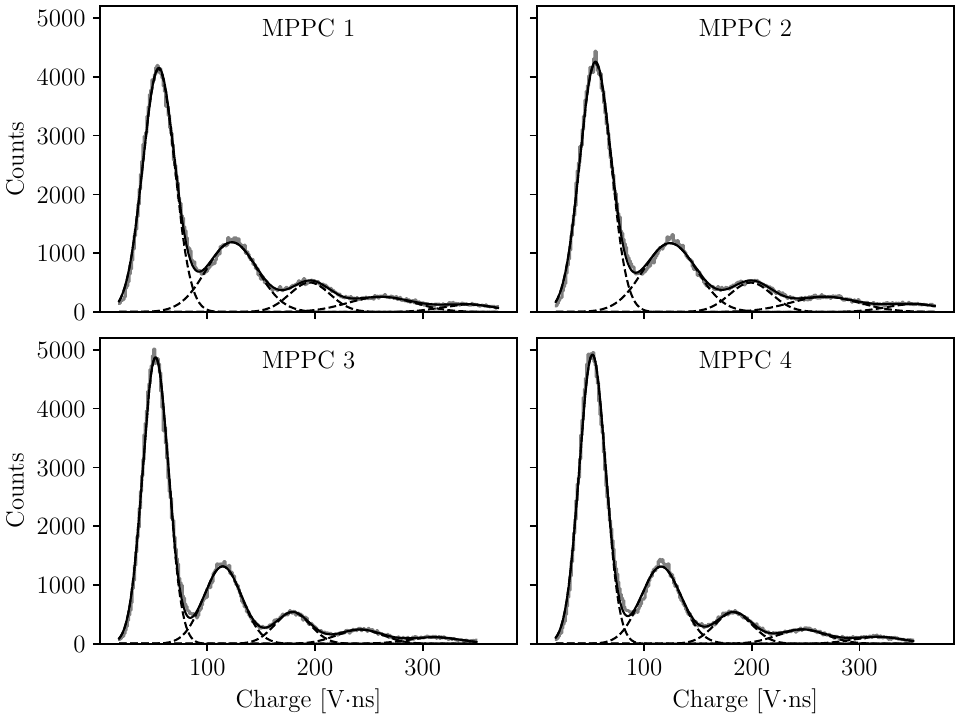}
    \caption{Single-photoelectron calibration of the four SiPMs at a temperature of 25~$^{\circ}$C. For each of the four channels, the single-photoelectron charge spectrum obtained from dark-pulse triggering is shown. The spectra are fitted with a sum of five Gaussian functions (solid black line) to extract the single-photoelectron response for gain calibration at the corresponding operating temperature.}
    \label{fig:single_pe}
\end{figure}

The resulting charge spectra are then used to extract the single photoelectron response, which serves as the reference calibration for relating the integrated charge to the number of photoelectrons under the corresponding temperature and operating conditions. Single-photoelectron spectra acquired with the SiPM at room temperature are shown in fig.~\ref{fig:single_pe}. 

\begin{figure}[ht!]
    \centering
    \includegraphics[width=1\linewidth]{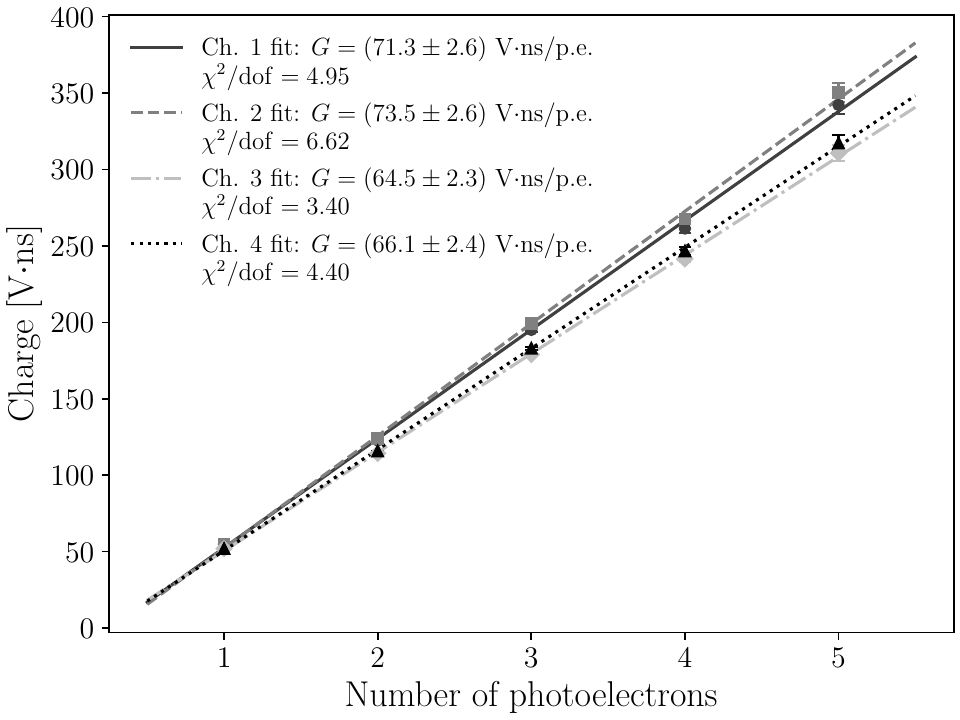}
    \caption{Linearity of the MPPC charge response for a run at temperature of 25~$^{\circ}$C. The mean integrated charge associated with the photoelectron peaks, extracted from the single-photoelectron spectra, is shown as a function of the number of photoelectrons. A linear fit is overlaid to verify the proportionality of the response $G$ in the single-photoelectron regime. The reported $\chi^2/\mathrm{dof}$ values are statistical only.}
    \label{fig:linearity}
\end{figure}

The linearity of the response was verified by comparing the charge associated with the photoelectron peaks and fitting a linear function, as shown in fig.~\ref{fig:linearity}. The relative gain variation among the four SiPM channels was found to be approximately 5\% RMS. The uncertainty associated with the SiPM gain calibration, results in an average relative uncertainty of 3.6\% on the reconstructed charge and is treated as a systematic uncertainty for the calculation of the light collection efficiency in section~\ref{sec:alpha}.

\section{Results}\label{sec:results}
\subsection{Detection of alpha particles with the SiPM setup}\label{sec:alpha}
\begin{figure}[ht!]
    \centering
    \includegraphics[width=1\linewidth]{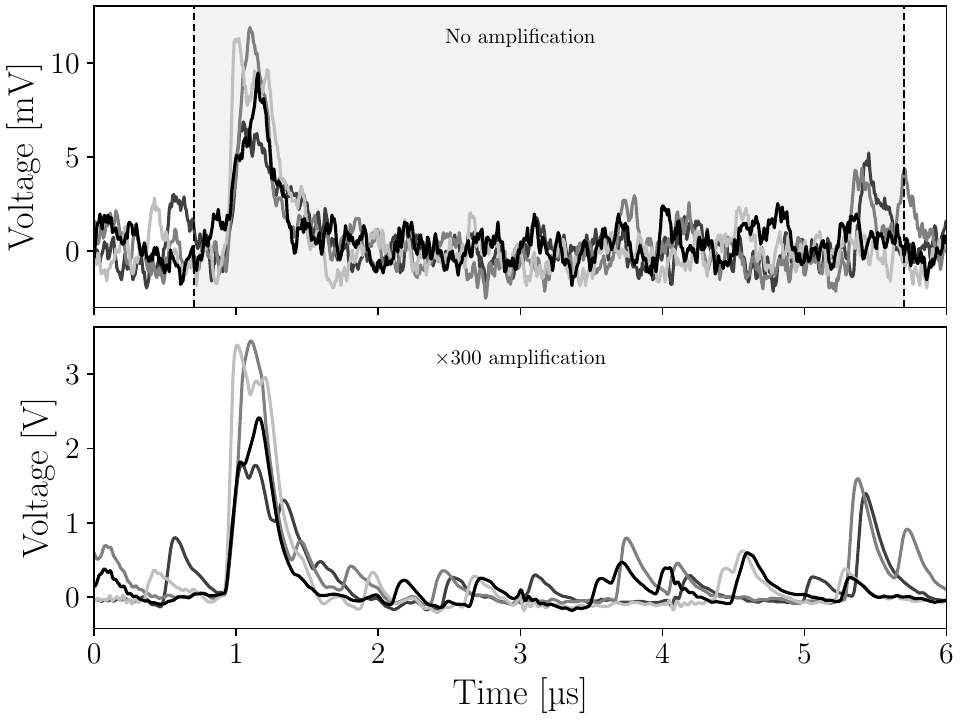}
    \caption{Waveforms acquired by the CAEN DT2740 digitizer for an event in gaseous xenon at 8.5 bar. Top panel: non-amplified waveforms with the 5~\si{\micro\second} integration window in gray. Bottom panel: waveforms after the $\times300$ voltage amplification performed by the front-end board. }
    \label{fig:waveforms}
\end{figure}

The data acquired with the \(\mathrm{^{241}Am}\) source are used to estimate the light collection efficiency (LCE) of the detector. Low-gain waveforms for the four channels were acquired with a majority-3 trigger on the high-gain waveforms set at approximately 2 photoelectrons. Such low trigger threshold was selected to enable the detection of 59.5~keV gamma emissions from $^{241}$Am decays, as detailed in section~\ref{sec:gamma}. Charge spectra were obtained by integrating the recorded waveforms over a 5~\si{\micro\second} time window for xenon (shown in fig.~\ref{fig:waveforms}) and 12~\si{\micro\second} time window for argon, given the presence of delayed recombination-induced scintillation (see section~\ref{sec:taus} for further discussion). 

\begin{figure*}[ht!]
    \centering
        \includegraphics[width=1\linewidth]{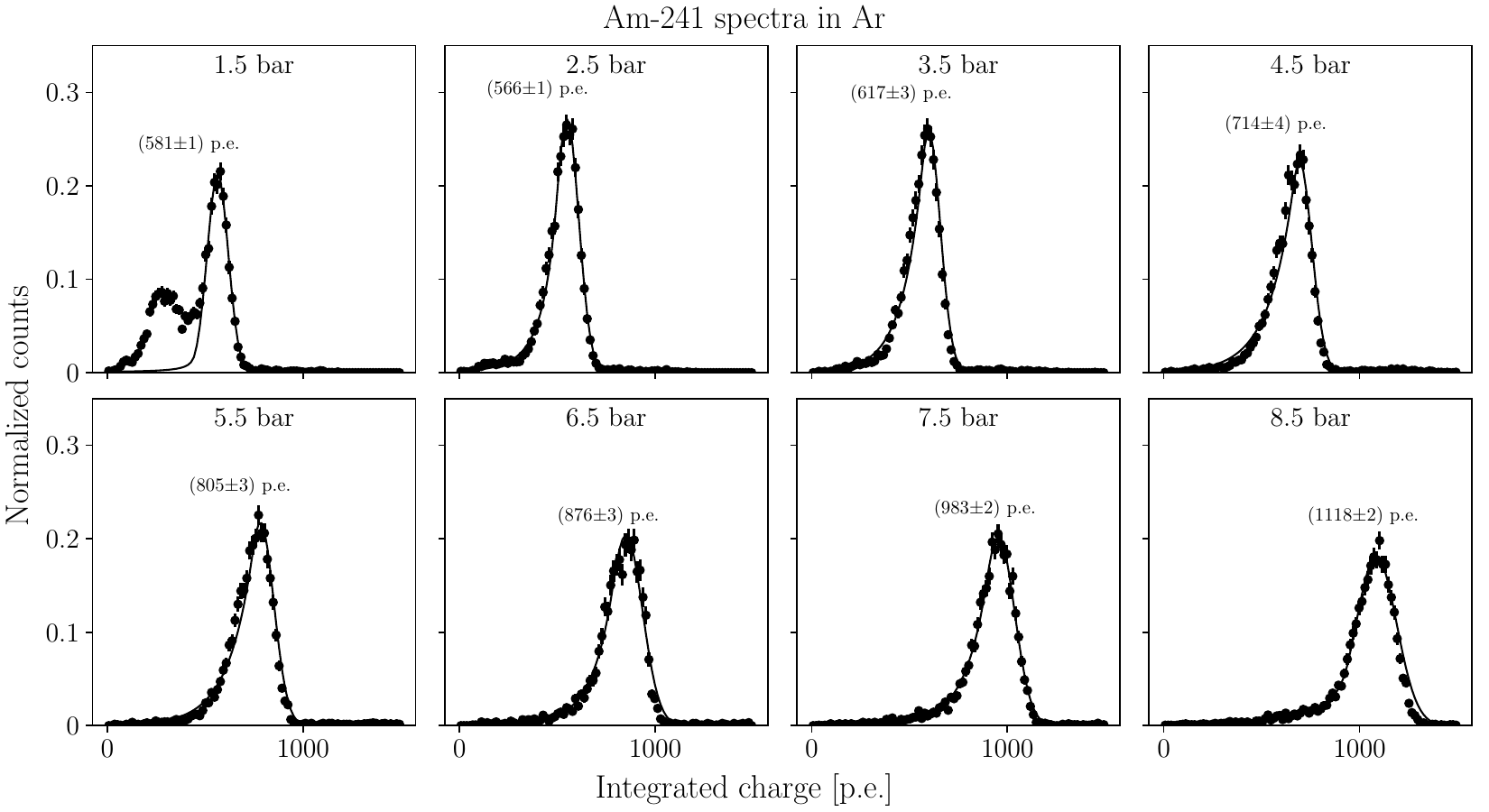}\vspace{2em}
    \includegraphics[width=1\linewidth]{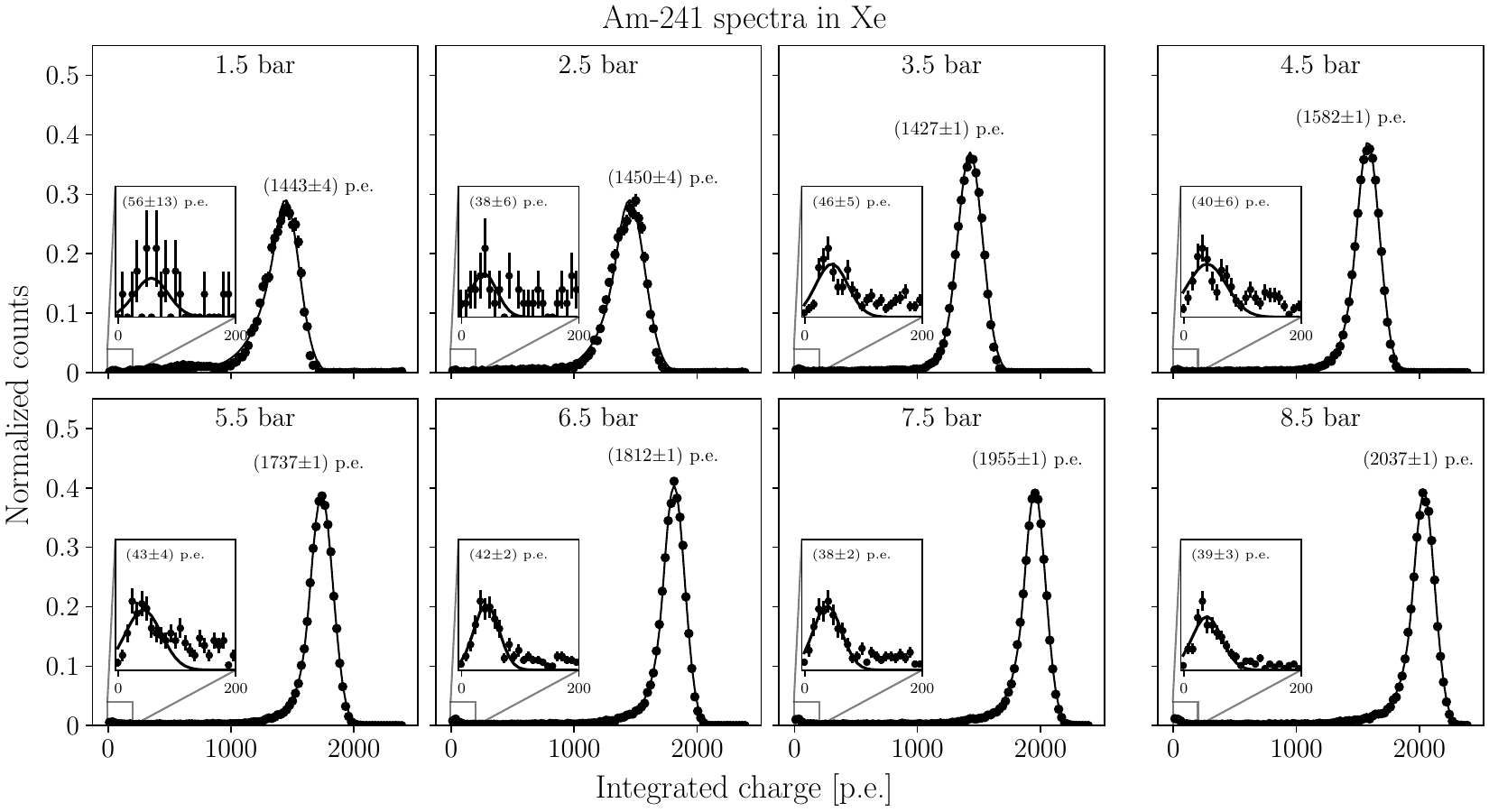}
    \caption{Integrated charge spectra obtained with the \(\mathrm{^{241}Am}\) source at different argon (top) and xenon (bottom) pressures between 1.5 and 8.5~bar. The spectra are shown after conversion to photoelectrons using the SiPM calibration and are normalized to unit area. The number of detected photoelectrons for the alpha peak, extracted from a Crystal Ball fit~\cite{Gaiser:1982yw} to each distribution (solid line), is indicated in each panel. The low-energy peak of the argon dataset at 1.5~bar is caused by a fraction of alpha rays reaching the side panels before depositing their full energy in the gas. In the case of the xenon data, the insets show the region between 0 and 200 p.e., which corresponds mainly to 59.5~keV gamma rays interacting in the active volume. The distributions have been fitted with a Gaussian, whose central value is indicated on top.}
    \label{fig:alphas}
\end{figure*}

\begin{figure*}[ht!]
    \centering
    \includegraphics[width=0.48\linewidth]{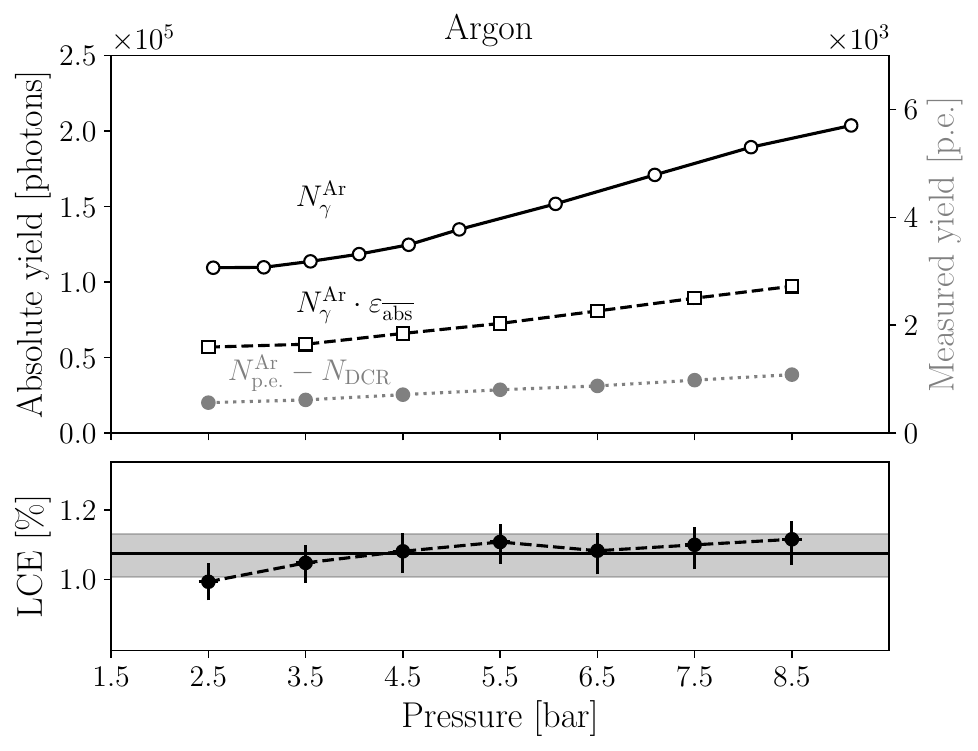}\qquad
    \includegraphics[width=0.48\linewidth]{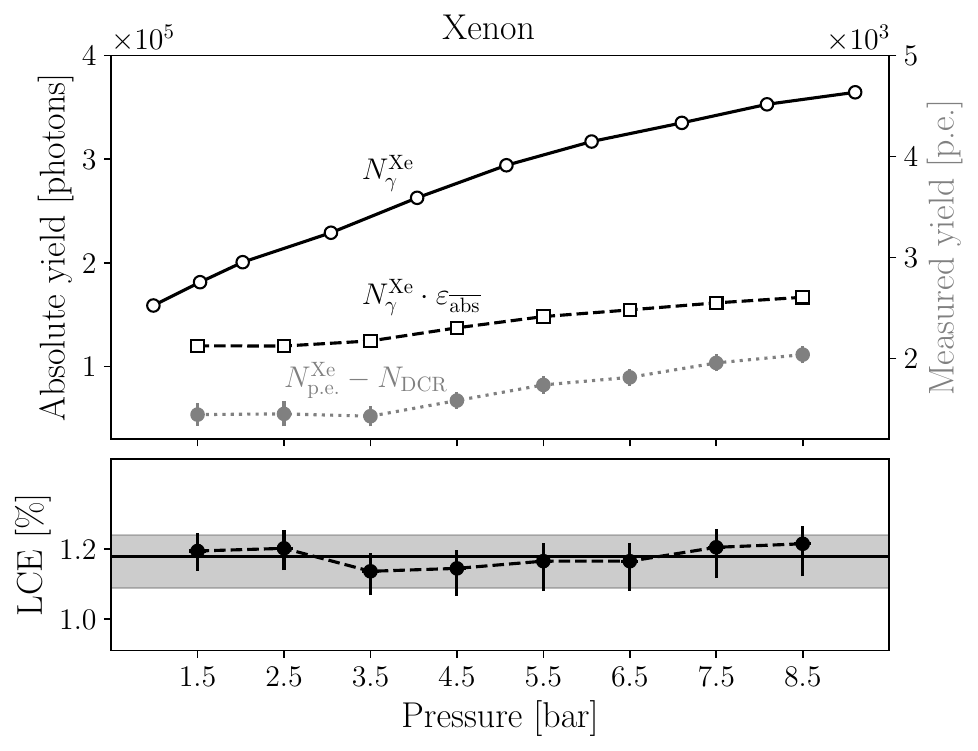}
    \caption{Upper panel: scintillation light yield obtained with the \(\mathrm{^{241}Am}\) source as a function of argon (left) and xenon (right) pressure. The measured light yield in photoelectrons minus the dark rate contribution $N_{p.e.} - N_{\mathrm{DCR}}$ (gray, right axis) is compared to the expected absolute scintillation photon yield $N_{\gamma}$ from ref.~\cite{Saito:2003dz} and its geometry-corrected, linearly interpolated value $N_{\gamma}\cdot\varepsilon_{\mathrm{\overline{abs}}}$ (black, left axis).  Lower panel: the efficiency, obtained with eq.~\eqref{eq:lce}, is found to be consistent with a constant value over the explored pressure range. In the case of argon, no absolute yield is provided in ref.~\cite{Saito:2003dz} for pressures below 2~bar, so the efficiency has not been calculated for the 1.5~bar data point. The solid line indicates the average value and the filled gray area corresponds to the uncertainty, giving $\mathrm{LCE}_{\mathrm{Xe}} = 1.18 \pm 0.01~\mathrm{(sta.)}~^{+0.07}_{-0.09}~(\mathrm{sys})~\%$ and $\mathrm{LCE}_{\mathrm{Ar}} = 1.07 \pm 0.01~\mathrm{(sta.)}~^{+0.06}_{-0.08}~(\mathrm{sys})~\%$, respectively.}
    \label{fig:lce}
\end{figure*}

The peaks corresponding to the 5.49 MeV alpha, shown in fig.~\ref{fig:alphas}, are compared in fig.~\ref{fig:lce} to the expected number of scintillation photons. The absolute scintillation yield for alphas was taken from values reported in ref.~\cite{Saito:2003dz} as a function of pressure.

As noted above, in the present configuration, no electric field is applied across the active volume. As a consequence, ionization electrons recombine locally, leading to an additional contribution to the scintillation signal from the recombination process. This effect enhances the total light yield with respect to the field-on case and increases with pressure~\cite{Saito:2003dz, NEXT:2012vuv}, as discussed in section~\ref{sec:discussion}. This contribution is taken into account when relating the measured charge spectra to the absolute scintillation photon yield, allowing the LCE of the detector to be extracted as a function of pressure.

\begin{figure*}
    \centering
    \includegraphics[width=0.9\linewidth]{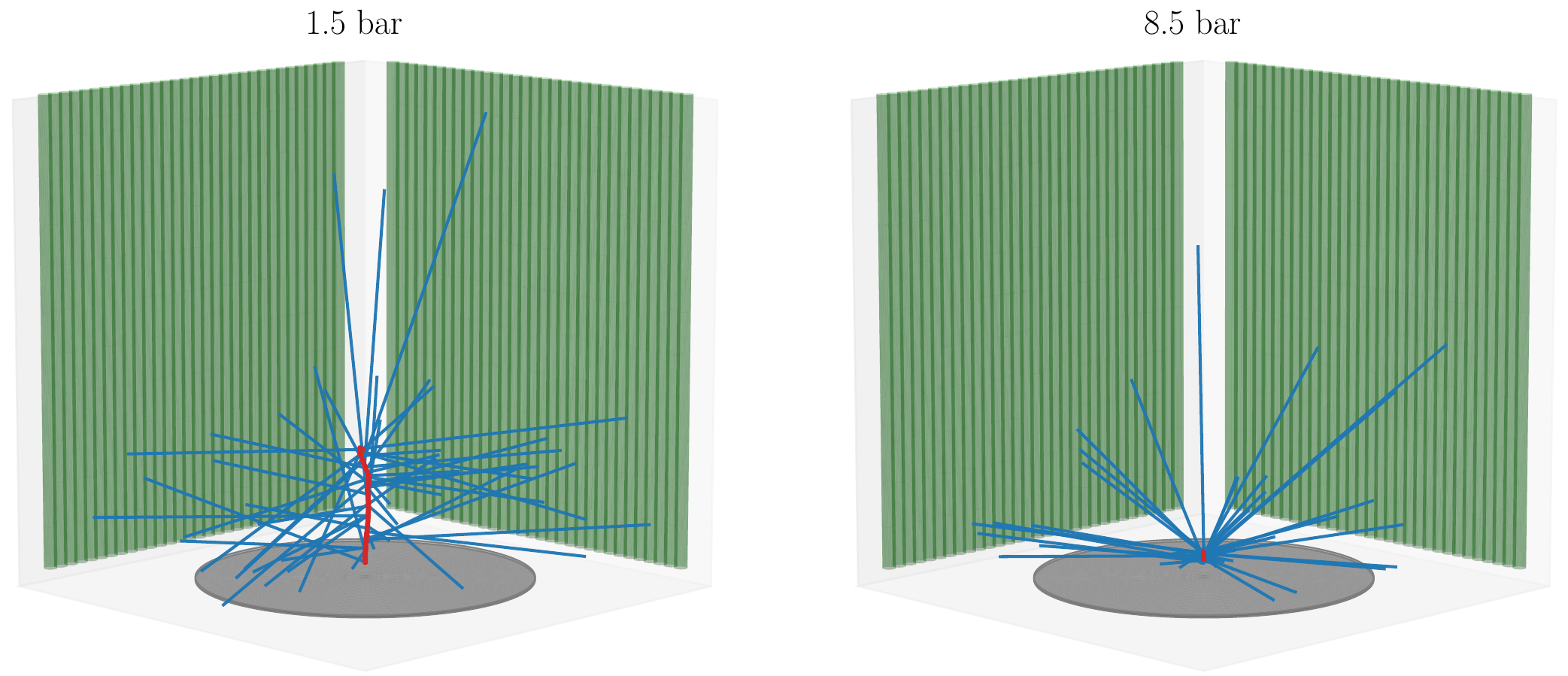}
    \caption{Geant4 simulation of $\alpha$ particle interactions in gaseous xenon at 1.5 bar (left) and 8.5 bar (right) inside the WLS fiber detector. Red lines show $\alpha$ particle tracks from the $\mathrm{^{241}Am}$ source deposited on a stainless-steel disk (gray), while a subset of scintillation photons is shown in blue. At low pressure, the longer $\alpha$ particle range shifts the scintillation region away from the disk, reducing photon absorption by the disk. At high pressure, the shorter range localizes scintillation near the disk surface, increasing absorption.}
    \label{fig:shadowing}
\end{figure*}

In addition, the determination of the light collection efficiency must account for geometrical and material effects specific to the experimental configuration. In the SiPM setup the \(\mathrm{^{241}Am}\) source, with an activity of 2.9~kBq, is deposited at the center of a stainless-steel disk with a diameter of 25~mm, which introduces a strong light absorption effect due to the very low reflectivity of stainless steel in the VUV. This light absorption effect is pressure dependent, as the range of alpha particles scales inversely with gas density, effectively displacing the scintillation region away from the source disk and reducing the solid angle subtended by the steel support, as exemplified in fig.~\ref{fig:shadowing}. 
Furthermore, the PTFE endcaps of the detector are not coated with TPB, leading to an additional, second-order loss of scintillation photons due to absorption at the PTFE surfaces, whose reflectivity in the VUV range is at most 66\%, depending on the manufacturing process~\cite{Silva:2009ip}.

Both effects reduce the number of scintillation photons reaching the fibers and therefore bias the raw estimation of the LCE. Corrections for these effects were applied, with their magnitude estimated using a dedicated Monte Carlo simulation based on \textsc{Geant4}~\cite{GEANT4:2002zbu}. In the baseline configuration, stainless steel was assumed to be fully absorbing, while PTFE reflectivities were implemented with wavelength-dependent values taken from ref.~\cite{Silva:2009ip}. Optical surface interactions were modeled using the \texttt{GLISUR} model \cite{023RaPC21211062K}.

\begin{figure}
    \centering
    \includegraphics[width=1\linewidth]{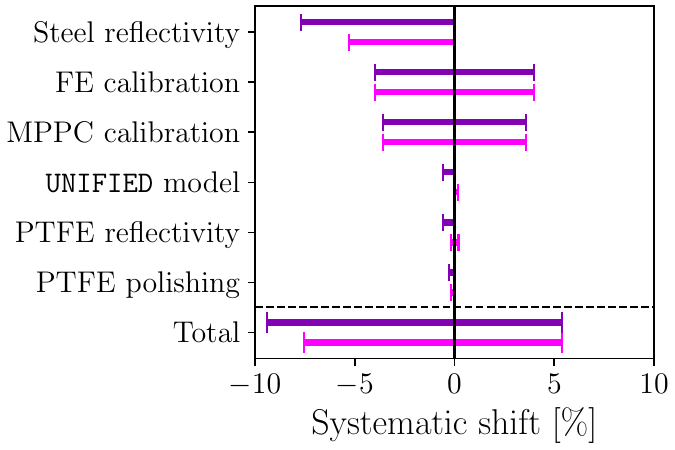}
    \caption{Relative systematic shifts on the LCE arising from variations in the optical model used in the \textsc{Geant4} simulation and from the calibration procedures for xenon (purple) and argon (fuchsia) at 8.5~bar. The total systematic uncertainty, obtained by combining all contributions, is also reported.}
    \label{fig:sys}
\end{figure}

The associated systematic uncertainty, listed in tab.~\ref{tab:detectable}, was evaluated by varying the optical properties and surface parameters within reasonable bounds. These variations included changes to the PTFE reflectivity in the visible and VUV ranges, the introduction of non-zero reflectivity for stainless steel, modifications to the surface polish parameters, and the use of the \texttt{UNIFIED} surface model~\cite{591410} as an alternative to \texttt{GLISUR}. Among these, the assumed reflectivity of the stainless steel surfaces constitutes the dominant source of systematic uncertainty. A breakdown of the relative impact of each component is shown in fig.~\ref{fig:sys} for the 1.5~bar data point.

In addition, due to the relatively large integration time, a non-negligible amount of dark pulses is generated by the SiPMs, which has been estimated with dedicated runs at the corresponding gain and temperature conditions.

Thus, the LCE at different pressures is estimated with the following equation: 
\begin{equation}\label{eq:lce}
\mathrm{LCE} = \frac{N_{\mathrm{p.e.}} - N_{\mathrm{DCR}}}{N_{\gamma}\cdot\varepsilon_{\mathrm{\overline{abs}}}},
\end{equation}
where $N_{\mathrm{p.e.}}$ is the number of photoelectrons corresponding to the $\alpha$ peak, obtained with a Crystal Ball fit~\cite{Gaiser:1982yw} of the charge spectra (shown in fig.~\ref{fig:alphas}), $N_{\mathrm{DCR}}$ is the number of photoelectrons generated by the dark noise and estimated with dedicated runs, $N_{\gamma}$ is the number of absolute scintillation photons produced by the $\alpha$ particle, obtained from ref.~\cite{Saito:2003dz}, and $\varepsilon_{\mathrm{\overline{abs}}}$ is the fraction of scintillation photons \emph{not} absorbed by the disk or by the endcaps, estimated by the simulation. The values of $\varepsilon_{\mathrm{\overline{abs}}}$ and $N_{\mathrm{DCR}}$ for both xenon and argon are available in tab.~\ref{tab:detectable} with the relative uncertainties.

\begin{table}
\begin{center}

\caption{Fraction of scintillation photons $\varepsilon$ \emph{not} absorbed by the stainless steel $^{241}$Am source or by the PTFE endcaps and number of photoelectrons generated by the dark noise in the same integration window as a function of pressure. Values are shown separately for argon and xenon and are used to obtain the LCE from eq.~\eqref{eq:lce}.}
\label{tab:detectable}
\setlength{\tabcolsep}{4pt} 
\small

\resizebox{\columnwidth}{!}{\begin{tabular}{c cc cc}
\hline\noalign{\smallskip}
 & \multicolumn{2}{c}{Argon} & \multicolumn{2}{c}{Xenon} \\
Pressure\\(bar) 
& $\varepsilon_{\mathrm{\overline{abs}}}$ 
& $N_{\mathrm{DCR}}$~(p.e.)
& $\varepsilon_{\mathrm{\overline{abs}}}$ 
& $N_{\mathrm{DCR}}$~(p.e.) \\
\noalign{\smallskip}\hline\noalign{\smallskip}
1.5 & $0.65^{+0.04}_{-0.05}$ & $24.5\pm0.1$ & $0.57^{+0.03}_{-0.04}$ & $12.6\pm0.1$ \\
2.5 & $0.58^{+0.03}_{-0.04}$ & $21.0\pm0.1$ & $0.53^{+0.03}_{-0.04}$ & $11.3\pm0.1$ \\
3.5 & $0.54^{+0.03}_{-0.04}$ & $20.8\pm0.1$ & $0.51^{+0.03}_{-0.04}$ & $10.8\pm0.1$ \\
4.5 & $0.53^{+0.03}_{-0.04}$ & $19.5\pm0.1$ & $0.49^{+0.03}_{-0.04}$ & $10.3\pm0.1$ \\
5.5 & $0.51^{+0.03}_{-0.04}$ & $18.3\pm0.1$ & $0.49^{+0.03}_{-0.04}$ & $10.1\pm0.1$ \\
6.5 & $0.51^{+0.03}_{-0.04}$ & $14.1\pm0.1$ & $0.48^{+0.03}_{-0.04}$ & $9.7\pm0.1$ \\
7.5 & $0.50^{+0.03}_{-0.04}$ & $15.9\pm0.1$ & $0.47^{+0.03}_{-0.04}$ & $9.4\pm0.1$ \\
8.5 & $0.49^{+0.03}_{-0.04}$ & $16.7\pm0.1$ & $0.47^{+0.03}_{-0.04}$ & $9.1\pm0.1$ \\
\noalign{\smallskip}\hline
\end{tabular}}
\end{center}
\end{table}




The resulting LCEs, shown in fig.~\ref{fig:lce} as a function of pressure for both gases, are found to be consistent with the following constant values:
\begin{align}
\mathrm{LCE}_{\mathrm{Xe}} = 1.18 \pm 0.01~\mathrm{(sta.)}~^{+0.07}_{-0.09}~(\mathrm{sys})~\%~ \\
\mathrm{LCE}_{\mathrm{Ar}} = 1.07 \pm 0.01~\mathrm{(sta.)}~^{+0.06}_{-0.08}~(\mathrm{sys})~\%, \nonumber
\end{align}
respectively for xenon and argon. The systematic uncertainty include contributions from the Monte Carlo model used to evaluate $\varepsilon_{\mathrm{\overline{abs}}}$ (fig.~\ref{fig:sys}), SiPM and front-end calibration, and dark noise subtraction. Importantly, this definition of LCE means the reported values correspond only to photons emitted or reflected toward the fibers.

These values can be compared with a back-of-the-envelope estimate based on parameters available in the literature and on dedicated Monte Carlo simulations. Following the treatment in ref.~\cite{GERDA:2022hxs}, the light collection efficiency can be expressed as the product of several contributions associated with wavelength shifting, photon transport, and photosensor response.

The trapping efficiency of the multi-clad Kuraray fibers is one of the dominant contributions. The manufacturer quotes a value of $\varepsilon_{\mathrm{trap}} \approx 5.4\%$~\cite{Kuraray}, consistent with the classical meridional-ray approximation. Analytical treatments that additionally account for skew rays predict a total trapping efficiency approaching $\sim 11\%$ for fibers with similar refractive indices~\cite{Achenbach:2003xlq,weiss}. The TPB absorption and re-emission efficiency for VUV photons is taken to be $\varepsilon_{\mathrm{TPB}} \approx 60\%$~\cite{Benson:2017vbw}, accounting also for the wavelength-shifted photons emitted towards the other fiber walls. The spectral overlap, defined as the integral of the product of the normalized TPB emission spectrum and the normalized Y11 absorption spectrum over wavelength, is estimated as $\varepsilon_{\mathrm{WLS}} \approx 85\%$. The photon detection efficiency of the SiPM is $\mathrm{PDE} \approx 46\%$ for the Y11 emission spectrum~\cite{Hamamatsu} (see fig.~\ref{fig:emission}). Additional factors include the presence of an aluminum mirror on one end of the fibers, corresponding to an effective $\varepsilon_{\mathrm{Al}} \approx 87.5\%$~\cite{Saraiva:2004cn}, propagation losses along the fiber ($\varepsilon_{\mathrm{att}} \approx 70\%$ for short distances~\cite{Kuraray, Bae:2025cww, Isabel:2026ixg}), and the optical coupling between the fiber and the photosensor, which is estimated to be $\varepsilon_{\mathrm{coup}} \approx 80\%$.

A simple product of these terms suggests a light collection efficiency around the percent level. In addition, in our setup, photons that are not trapped on first incidence can be reflected by the PTFE panels and endcaps and subsequently re-intercept the fibers, effectively increasing the probability of eventual detection. 

Therefore, compared to ref.~\cite{GERDA:2022hxs}, the higher LCE observed in this work can be attributed to the four main factors: (1) negligible absorption by impurities in few cm of GXe (while contributing by an $e$ factor in LAr), (2) improved spectral matching of Y11 fibers with TPB emission compared with the BCF-91A ones, (3) higher PDE of the Hamamatsu MPPCs (46\% vs. 30\%), and (4) presence of a PTFE-lined geometry that enhances photon recycling. The modestly higher LCE observed in xenon relative to argon in our setup can be attributed to the increased $\varepsilon_{\mathrm{TPB}}$ at 175 nm compared to 128 nm~\cite{Benson:2017vbw}.

\subsection{Detection of alpha particles with the PMT setup}
\label{sec:frogxe}
An independent measurement of the light collection efficiency with alpha particles was performed in the setup instrumented with red-enhanced PMTs and WLS fibers from a different manufacturer. The data collected with this detector corresponds to alpha interactions in the xenon volume. Although the VUV PMT was originally intended to provide both the event trigger and an independent measurement of the light yield, it was not operational during this data-taking period. Instead, data acquisition was performed by triggering on one of the red-enhanced PMTs. To mitigate potential trigger-induced biases, the dataset combines four independent runs, each using a different PMT as the trigger.

\begin{figure}
    \centering
    \includegraphics[width=\columnwidth]{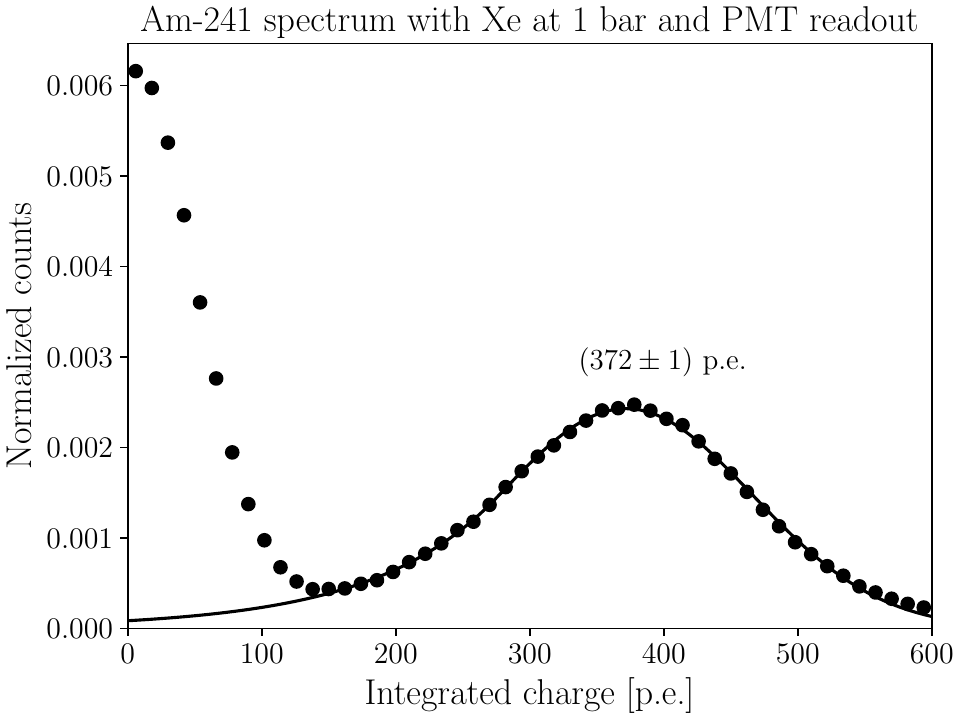}
    \caption{Integrated PMT charge spectrum for the xenon run at 1 bar. The spectrum includes a conversion to photoelectrons by a dedicated single-photoelectron calibration. The spectrum features a main distribution at $\sim$372~p.e. attributed to alpha particles depositing most of their energy in the gas, and a low-energy population interpreted as partial energy depositions in the source cover overlapping the 59.5-keV gamma-ray peak. The number of detected photoelectrons for alpha interactions is extracted from a Crystal Ball fit to the high energy population.}
    \label{fig:frogxe-spectrum}
\end{figure}

The spectrum of the integrated charge is shown in fig.~\ref{fig:frogxe-spectrum}. The spectrum exhibits a prominent peak at approximately 370 photoelectrons, attributed to $\alpha$ particles depositing virtually all their 4.6 MeV energy within the gas volume. A lower-energy population is also observed, attributed to alpha particles losing part of their energy in the source cover, overlapping a smaller contribution consistent with a combination of 59.5 keV $\gamma$ interactions (see section~\ref{sec:gamma}).

As in section \ref{sec:alpha}, we base our estimate of the light collection efficiency on Eq.~\ref{eq:lce}. In this case, a Crystal Ball fit to the high-energy population yields $N_{\mathrm{p.e.}} = (372\pm1)$~p.e., while the contribution from the dark count rate ($N_{\mathrm{DCR}}$) is negligible ($N_{\mathrm{DCR}} \ll 1$). The number of emitted scintillation photons, $N_\gamma$, is computed following the same reference \cite{Saito:2003dz} giving 1.3$\times$10$^5$ photons at 4.6 MeV. 

The parameter $\varepsilon_{\mathrm{\overline{abs}}}$ is estimated from a dedicated optical simulation assuming the aforementioned reflectivities for PTFE and stainless steel. Considering further uncertainties associated with the reflectivity of the VUV-PMT window and of  Vikuiti at 172 nm, the simulation yields $\varepsilon_{\overline{\mathrm{abs}}}=0.64\pm0.03$, from which we obtain $\textrm{LCE}_{\textrm{Xe}}^{\textrm{PMT}} = 0.45\pm 0.01 \textrm{ (sta.)} \pm 0.05 \textrm{ (sys.)}$~\%.

To compare this result with measurements performed using SiPMs, the difference in photon detection efficiency (PDE) must be taken into account. The PMTs have an average PDE of 18.4\% over the fiber emission spectrum, whereas the average SiPM PDE is 46\%. Applying a scaling factor of 46/18.4, the extrapolated LCE value becomes ${\textrm{LCE}_{\textrm{Xe}}^{\textrm{SiPM (extr)}} = 1.12 \pm 0.01 \textrm{ (sta.)} \pm 0.12 \textrm{ (sys)}~\%}$. This value is in good agreement with the corresponding SiPM measurements. However, direct comparison is complicated by differences in the specifications provided by each manufacturer. The Y11 fibers have both a higher trapping efficiency than the BCF-91A fibers (5.4\% versus 3.44\% in the meridional approximation) and a better overlap between the TPB emission spectrum and the fiber absorption spectrum. These factors would suggest a higher intrinsic LCE for the setup with the Y11 fibers. The fact that the BCF-91A fibers nevertheless appear competitive may indicate that the optical coupling between the fibers and photosensors was better in the PMT setup.

\subsection{Detection of 59.5~keV gamma rays}\label{sec:gamma}
In the present setup, the large active volume combined with the high pressure of the gas results in a non-negligible probability for 59.5~keV gamma rays to interact in the xenon and produce scintillation light. This feature is particularly relevant in view of low-energy signals such as the 41.5~keV decays of $\mathrm{^{83m}Kr}$, used for the calibration of the NEXT detectors~\cite{NEXT:2018sqd, NEXT:2025fpq}, demonstrating that the detector is sensitive not only to heavily ionizing alpha particles but also to low-energy interactions.

These interactions manifest as a small peak at the lower end of the xenon spectra shown in the insets of fig.~\ref{fig:alphas} (bottom), corresponding to approximately 40~p.e., which becomes more pronounced at high pressure. In contrast, in argon the corresponding effect is expected to be negligible, as a consequence of the lower gas density, the smaller gamma-ray cross section at these energies, and the lower light yield, such as that no discernible low-energy peak is observed in fig.~\ref{fig:alphas} (top).

In principle, this feature could be exploited to extract an independent estimate of the  light yield with xenon using the gamma-induced events, which would correspond to roughly $(40 - 10)~\mathrm{p.e.} / 59.5~\mathrm{keV}\approx0.5~\mathrm{p.e./keV}$ (assuming 10 p.e. as the average dark noise contribution, as listed in tab.~\ref{tab:detectable}). This value must be compared with 0.26-0.36 p.e./keV for alpha particles. 

However, although producing a reasonable estimate, such an approach is affected by substantial systematic uncertainties. First, the ionization density and recombination dynamics associated with gamma-induced electron tracks differ from those of alpha particles, such that the recombination behavior measured in ref.~\cite{Saito:2003dz} cannot be directly applied. Second, due to K$_{\alpha}$ and K$_{\beta}$ emission from xenon~\cite{NEXT:2014yrq}, energy deposition may be multi-site, which further complicates the determination of the light propagation and the associated light absorption effects. 
Finally, because the amount of scintillation light produced by these events is relatively small, the reconstructed position of the gamma-induced peak is highly sensitive to the dark noise and the detection threshold, which is determined by the trigger configuration and thus introduces an additional, configuration-dependent systematic uncertainty.

Given these limitations of the gamma-induced event subset, complementary measurements were performed with xenon using tagged cosmic-ray muons, described below. These events do not suffer from source shadowing effects and produce substantially larger light signals, providing a more robust and reliable handle for the characterization of the light yield.

\subsection{Detection of cosmic muons}\label{sec:muons}

\begin{figure*}[ht!]
    \centering
    \includegraphics[width=1\linewidth]{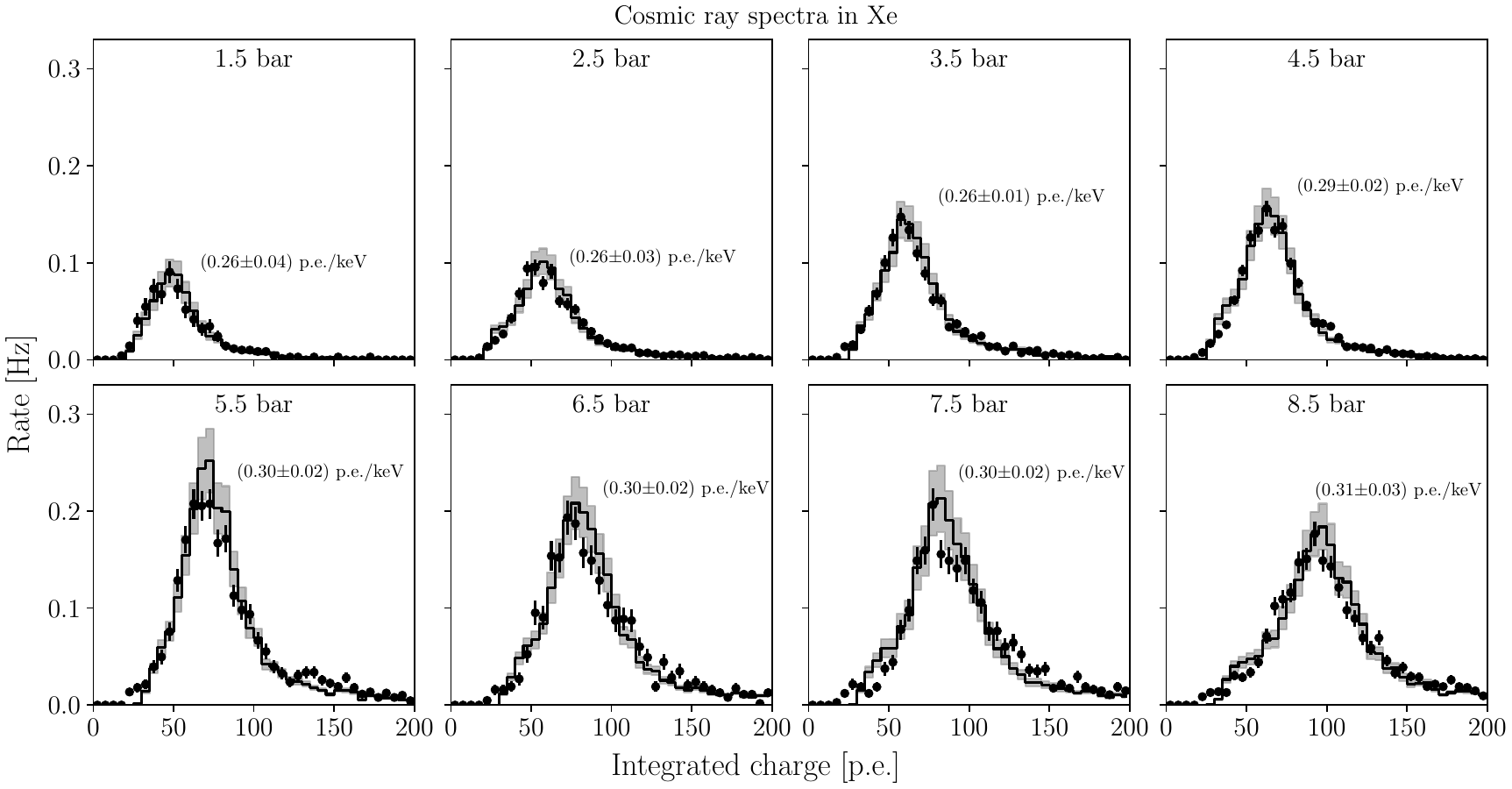}
    \caption{Integrated charge spectra for cosmic muons at different xenon pressures. The distributions show the reconstructed integrated charge for through-going muon events selected with the coincidence trigger and offline photoelectron requirement. The solid lines correspond to the best-fit comparison with \textsc{Geant4} Monte Carlo simulations, from which the light yield is extracted and the shaded band represents the $1\sigma$ model uncertainty from propagation of the scaling factor uncertainty and the uncertainty on the area normalization to data. The resulting light yields, expressed in p.e./keV, are indicated in each panel.}
    \label{fig:cosmic}
\end{figure*}

Cosmic muons are selected using a coincidence trigger between the two plastic scintillator panels placed above and below the horizontal pressure vessel. Because the scintillator panels are significantly larger than the active volume, an additional offline requirement of at least 3 p.e. per channel is imposed to suppress events not associated with energy deposition in the gas. This configuration preferentially selects through-going muons traversing the active volume, with an average track length of approximately 4~cm in the gas. The resulting charge spectra are obtained following the same reconstruction and calibration procedure described above, enabling a comparison with the results obtained using the \(\mathrm{^{241}Am}\) source.

It should be noted, however, that the comparison between the two data sets is not direct, since in the case of cosmic muons the WLS fibers are themselves crossed by ionizing radiation~\cite{Bae:2025cww}. WLS fibers, while not designed as scintillators, can nonetheless produce light when crossed by ionizing radiation as a result of the intrinsic scintillation of the polystyrene core~\cite{chakraborty2017ionoluminescence}. This contribution was quantified in a dedicated run under vacuum conditions and was found to be, on average, $(17\pm1)$~p.e. per event, after correcting for the dark noise. For predominantly vertical cosmic muons, the typical trajectory intersects approximately two fiber cores. A small fraction of muons traverse the detector at very shallow angles and graze the vertically mounted fiber panels, crossing a larger number of fibers and producing a small high-energy tail in the distribution. Neglecting this component, the measured signal corresponds to approximately 8~p.e. per fiber, in reasonable agreement with the value reported in ref.~\cite{Bae:2025cww} for $\beta$-particle interactions.

The energy spectra of cosmic interactions are influenced by several factors, including the angular distribution of the cosmic rays, the detector acceptance, and the solid angle intercepted by the two endcaps from every point along the cosmic path. Thus, instead of using an analytical formula for the energy distribution, we determined the light yield by comparing the measured spectra with a dedicated \textsc{Geant4} Monte Carlo simulation employing the same detector geometry used to estimate the geometrical acceptance in the $\alpha$ runs described in section~\ref{sec:alpha}. Cosmic-ray muons are generated according to the expected atmospheric angular distribution at sea level, allowing the simulation to reproduce the realistic spread of incidence angles and corresponding track lengths within the active volume. The simulated spectra include the intrinsic fiber scintillation component, as estimated above, and the contribution from the SiPM dark rate, measured with dedicated runs. In the simulation, the number of detected photoelectrons is smeared according to a Poisson distribution to account for the statistical nature of photon detection. The light yield is treated as a free parameter and is determined by minimizing the difference between the simulated and measured charge distributions with a $\chi^2$ test, as shown in fig.~\ref{fig:cosmic}. 

A slight increasing trend in the light yield with increasing pressure is observed, from $(0.26\pm0.04)$~p.e./keV at 1.5 bar to $(0.31\pm0.03)$~p.e./keV at 8.5 bar, following the same trend observed for alpha particles, albeit with a smaller relative variation (see fig.~\ref{fig:alphas}, bottom). Minimum-ionizing particles (MIPs) such as cosmic muons are expected to exhibit negligible recombination~\cite{Leardini:2021qnf}, leading to a light yield that is essentially independent of pressure. Thus, the slight increase observed here may arise from a combination of effects: (1) low-energy $\delta$ electrons producing locally denser ionization than the primary muon track and therefore a non-zero recombination effect, (2) a higher probability for xenon X-ray interaction inside the gas for increasing pressure, and (3) a residual pressure-dependent trigger bias.

\subsection{Average scintillation waveforms and time structure}\label{sec:taus}

\begin{figure*}[ht!]
    \centering
    \includegraphics[width=1\linewidth]{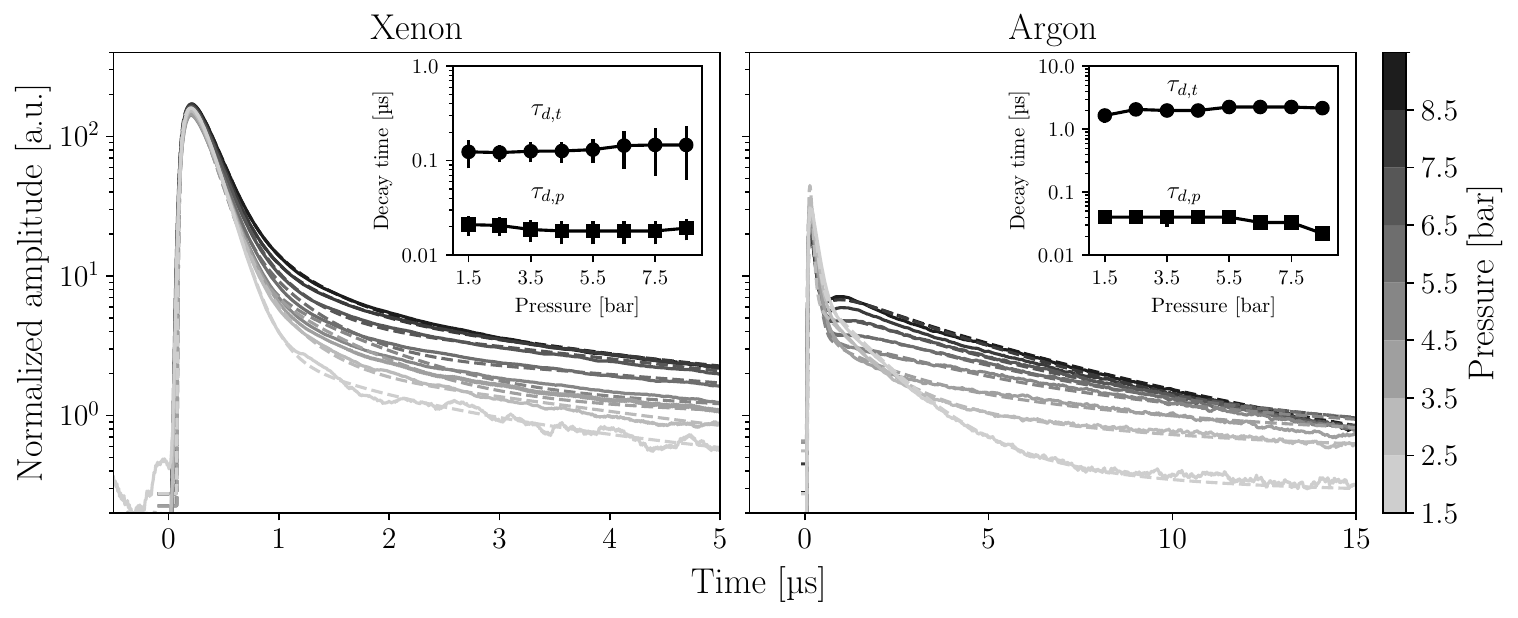}
    \caption{Average scintillation waveforms induced by Am-241 decays in gaseous xenon (left) and argon (right) at pressures between 1.5 and 8.5 bar. For each pressure, waveforms are time-aligned to the peak and averaged over many events. The color scale indicates the operating pressure. The average waveforms are fitted with the functions of eq.~\eqref{eq:tau_ar} and eq.~\eqref{eq:tau_xe} for argon and xenon, respectively. The insets show the decay time for the prompt component and the triplet component obtained from the fits as a function of pressure.}
    \label{fig:taus}
\end{figure*}

The time structure of the scintillation signal was investigated by constructing average waveforms from the alpha-induced events in gaseous xenon and argon at different operating pressures. For each pressure point, waveforms from the four SiPM channels were normalized to the same integrated charge and averaged to obtain a representative pulse shape.

The scintillation time profile in noble gases arises from several microscopic processes contributing on different timescales. In gaseous argon and xenon, three main components can be identified: a prompt contribution associated with the formation and decay of singlet excimers, a slower component originating from the decay of triplet excimers, and a delayed component produced by electron-ion recombination. In our setup, the prompt emission includes also the response of the TPB~\cite{Segreto:2014aia, Agnes:2021cqa} and the re-emission decay time of the WLS fibers~\cite{Brekhovskikh:2000pna}, both at the nanosecond level. The relatively slow intrinsic response of the SiPMs, of order $\sim80$~ns due to their large capacitance, prevents the direct resolution of the these processes from the singlet emission. 

The delayed emission at the microsecond scale originates from the recombination of thermalized electrons with molecular ions produced along the ionization track. Following the treatment in ref.~\cite{SUZUKI1982565}, the kinetics of this process can be described by the recombination equation
\begin{equation}
    \frac{dn}{dt}=-\alpha n^2,
\end{equation}
whose solution gives $n(t)=n_0/(1+t/\tau_r)$, where $\tau_r=(\alpha n_0)^{-1}$ is the characteristic recombination time, with $\alpha$ being the electron–ion recombination coefficient and $n_0$ the number density of thermalized electrons in the track. 

The relative importance of these processes depends on the gas and pressure. For argon at pressures below approximately $8\times10^3$ Torr ($\sim10.6$~bar), the characteristic recombination time is longer than the lifetime of the triplet excimer state~\cite{SUZUKI1982565}. This corresponds to our regime, where the delayed scintillation is governed primarily by recombination kinetics rather than by the excimer lifetime, leading to a $I(t)\propto(1+t/\tau_r)^{-1}$ tail in the time profile. 

Experimentally, the measured waveform corresponds to the convolution of the true scintillation emission profile with the impulse response of the SiPMs. For this reason, we fit the detected signal with the convolution of the SiPM response with the sum of three scintillation components: a prompt component representing the combined singlet, TPB, and fiber contributions, a triplet excimer component, and a delayed recombination component, as shown in fig.~\ref{fig:taus}. To improve the stability of the fit, the power-law time dependence of the recombination is approximated here with an effective exponential decay. The function used to describe the average waveforms is therefore:
\begin{align}\label{eq:tau_ar}
I_{\mathrm{Ar}}(t) = &\Bigg[
A_p \left(e^{-(t-t_0)/\tau_{d,p}} - e^{-(t-t_0)/\tau_{r,p}}\right)\\
&+ A_t \left(e^{-(t-t_0)/\tau_{d,t}} - e^{-(t-t_0)/\tau_{r,t}}\right)\nonumber\\
&+ A_r \left(e^{-(t-t_0)/\tau_{d,r}} - e^{-(t-t_0)/\tau_{r,r}}\right) \Bigg]\nonumber\\
& \otimes R_{\mathrm{SiPM}}(t)  + C ,\nonumber
\end{align}
where $R_{\mathrm{SiPM}}(t)$ represents the detector response, modeled as a difference of exponentials accounting for the finite rise and decay times of the photosensor, $A_i$ are the amplitudes of the three components ($p$ prompt, $t$ triplet, $r$ recombination, each one with its own rise $\tau_r$ and decay  $\tau_d$ time), and $C$ accounts for the constant baseline due to dark noise.

In xenon, a similar recombination mechanism occurs, although the different track structure and electron thermalization properties lead to a combination of an early exponential component, $r_1$ and a later power-law tail~\cite{SUZUKI1982565}, which we approximate also here with an effective exponential decay to improve the stability of the fit, $r_2$: 
\begin{align}\label{eq:tau_xe}
I_{\mathrm{Xe}}(t) = &\Bigg[
A_p \left(e^{-(t-t_0)/\tau_{d,p}} - e^{-(t-t_0)/\tau_{r,p}}\right)\\
&+ A_t \left(e^{-(t-t_0)/\tau_{d,t}} - e^{-(t-t_0)/\tau_{r,t}}\right)\nonumber\\
&+ A_{r_1} \left(e^{-(t-t_0)/\tau_{d,r_1}} - e^{-(t-t_0)/\tau_{r,r_1}}\right)\nonumber \\
&+ A_{r_2} \left(e^{-(t-t_0)/\tau_{d,r_2}} - e^{-(t-t_0)/\tau_{r,r_2}}\right) \Bigg]\nonumber \\
& \otimes R_{\mathrm{SiPM}}(t)\nonumber+ C.\nonumber
\end{align}

The prompt and triplet decay times obtained from the fits are approximately constant as a function of pressure, as shown in the insets of fig.~\ref{fig:taus}. Averaging over all pressures, we obtain $\tau_{d,t}^{\mathrm{Ar}} = (2.16 \pm 0.11)$~\si{\micro\second} and $\tau_{d,p}^{\mathrm{Ar}} = (39 \pm 5)$~ns for argon, and $\tau_{d,t}^{\mathrm{Xe}} = (121 \pm 12)$~ns and $\tau_{d,p}^{\mathrm{Xe}} = (19 \pm 3)$~ns for xenon. It is important to note that the prompt component reported here does not correspond to the intrinsic singlet lifetime of the noble gas, but represents an effective prompt time constant that includes, in addition, the convolution of the response of the TPB and the transport and re-emission times associated with the WLS fibers. Moreover, likely due to a strong correlation between the recombination and triplet components in argon obtained by the fit, the decay time of the latter was found to be lower than the typical literature values~\cite{Santorelli:2020fxn}.

\subsection{Estimation of $w_{sc}$}\label{sec:discussion}
\begin{figure}[ht!]
    \centering
    \includegraphics[width=1\linewidth]{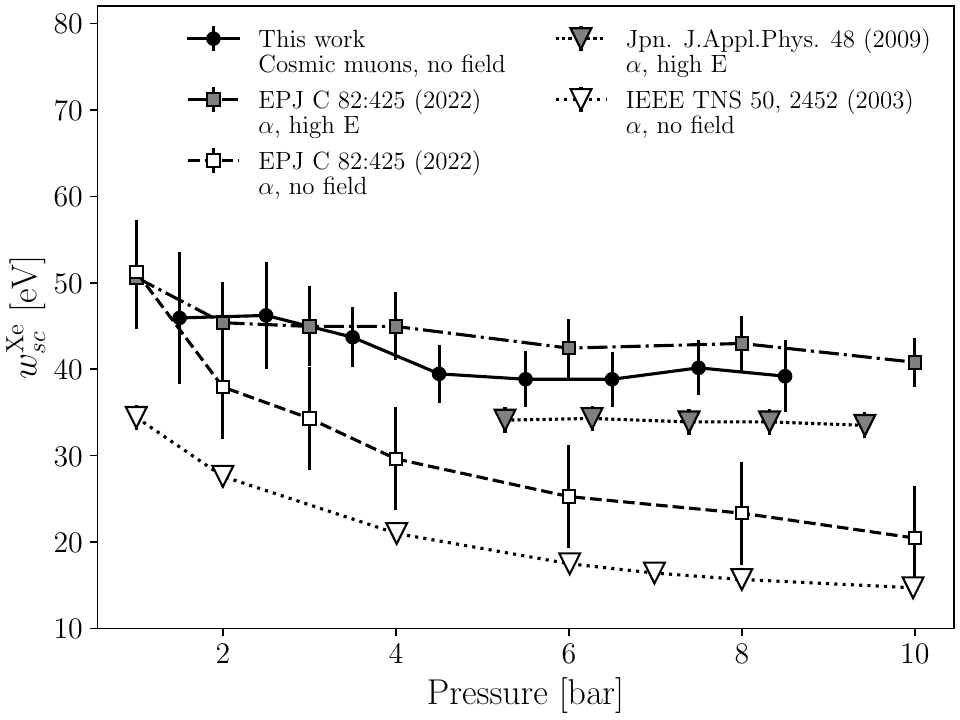}
    \caption{Average energy to produce a scintillation photon in xenon with cosmic rays (black dots), compared with 2$^{\mathrm{nd}}$ continuum data of ref.~\cite{Leardini:2021qnf} with and without electric field (gray and white squares) and band-unresolved data with (ref. \cite{mimura2009average}) and without electric field (ref.~\cite{Saito:2003dz}) (gray and white triangles).}
    \label{fig:w_sc}
\end{figure}

To quantify the impact of pressure on the scintillation yield, we compute the mean energy required to produce a scintillation photon in xenon at pressure $P$, $w_{\mathrm{sc}}^{\mathrm{Xe}}$, using:

\begin{equation}
w_{\mathrm{sc}}^{\mathrm{Xe}}(P) =
\frac{\mathrm{LCE}_{\mathrm{Xe}}}{\mathrm{LY}_{\mathrm{Xe}}(P)},\label{eq:w_sc}
\end{equation}
where $\mathrm{LY}_{\mathrm{Xe}}$ is the light yield obtained with the cosmic muon runs of section~\ref{sec:muons} (expressed in p.e./keV) and $\mathrm{LCE}_{\mathrm{Xe}}$ is the light collection efficiency, obtained in section~\ref{sec:alpha} (expressed in p.e./photon). No strong pressure dependence of $w_{\mathrm{sc}}^{\mathrm{Xe}}$, shown in fig.~\ref{fig:w_sc}, is observed, indicating that cosmic muons produce negligible recombination, even at high pressure, as expected for low-ionizing particles. The data agree within uncertainties both with the values of ref.~\cite{Leardini:2021qnf} in the presence of electric field, which was obtained by isolating only the 2$^{\mathrm{nd}}$ continuum emission, and with the band-unresolved data of ref.~\cite{mimura2009average}.

Considering only the 1.5~bar data point, eq.~\eqref{eq:w_sc} yields $w_{sc}^{\mathrm{Xe}} = 45\pm7~\mathrm{(sta.)}~^{+4}_{-5}~\mathrm{(sys.)}~\mathrm{eV}$. Because recombination is strongly suppressed for low-ionizing particles and especially in this low-pressure regime, this result effectively corresponds to the limit of vanishing recombination, analogous to xenon operated under a sufficiently strong drift field. Our measurement can be then compared with the systematic study of ref.~\cite{Henriques:2023evt} at 1.2~bar, which obtained $w_{\mathrm{sc}}^{\mathrm{Xe}} = 38.7 \pm 0.6~\mathrm{(sta.)}\,^{+7.7}_{-7.2}~\mathrm{(sys.)}\,\mathrm{eV}$ only assuming VUV emission and $w_{\mathrm{sc}}^{\mathrm{Xe}} = 43.5 \pm 0.7~\mathrm{(sta.)}\,^{+8.7}_{-8.1}~\mathrm{(sys.)}\,\mathrm{eV}$ isolating the VUV 2$^{\mathrm{nd}}$ continuum emission from the UV-VIS one of the 3$^{\mathrm{rd}}$ continuum. The uncertainty in our measurement does not allow to distinguish between the two scenarios.

\section{Discussion}
\begin{figure}[ht!]
    \centering
    \includegraphics[width=1\linewidth]{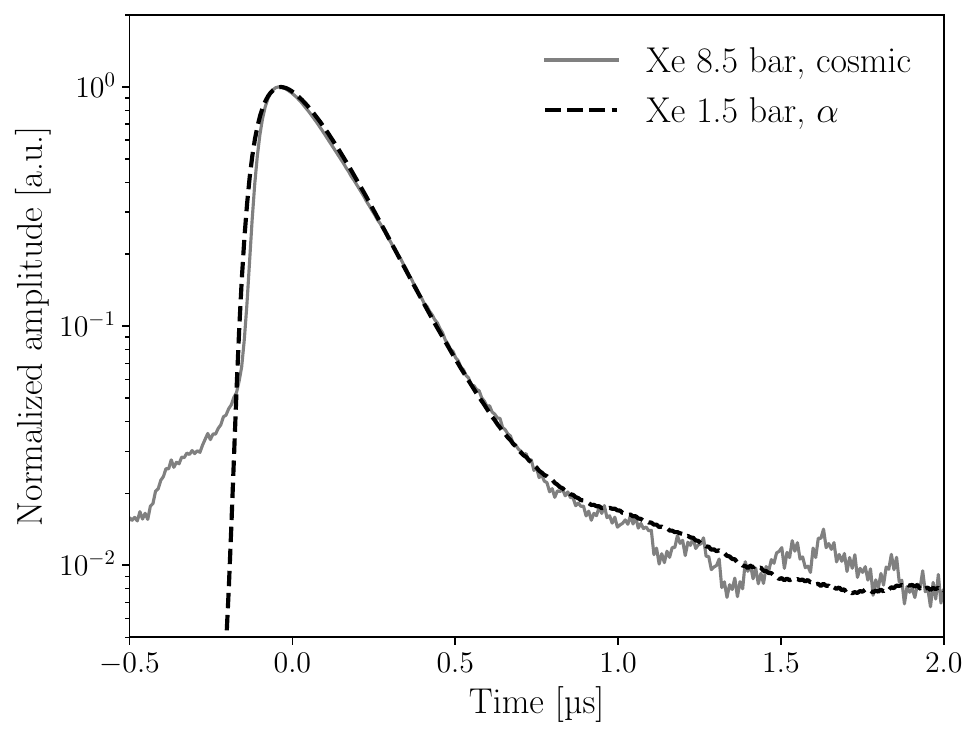}
    \caption{Time structure of a xenon run with $\alpha$ rays at low pressure (dashed black line), compared with a cosmic-ray run at high pressure (solid gray line).}
    \label{fig:alpha_vs_cosmic}
\end{figure}

To aid the interpretation of the observed pressure dependence of the light yield for alpha particles, it is useful to consider the role of electron–ion recombination in noble gases. At low pressure, recombination is expected to be strongly suppressed and the contribution of recombination-induced scintillation is negligible near atmospheric pressure~\cite{Saito:2003dz, Leardini:2021qnf}. In this regime, the observed light yield is therefore dominated by primary scintillation arising from direct excitation of the gas atoms. This expectation is also supported by our data: the average waveform measured at low pressure shows a strong suppression of the delayed component associated with recombination light both in xenon and argon (fig.~\ref{fig:taus}), indicating that recombination plays only a minor role under these conditions. In addition, the cosmic-ray muons used in this study behave as MIPs, producing a low ionization density along their tracks. Under such conditions recombination is expected to be negligible, as observed in previous measurements with $\beta$ particles~\cite{Leardini:2021qnf} and X-rays~\cite{Balan:2010kx}. The time structure of the $\alpha$ run at 1.5~bar exhibits a near perfect overlap with the cosmic run at 8.5~bar (see fig.~\ref{fig:alpha_vs_cosmic}), supporting this hypothesis. 

This situation is closely analogous to detector operation in the presence of an external electric field, as in a TPC, where the applied field efficiently separates ionization electrons from ions and suppresses recombination~\cite{Saito:2003dz, Leardini:2021qnf, Santorelli:2020fxn, Henriques:2023evt}. Consequently, measurements performed with cosmic muons and/or at sufficiently low pressure can be regarded as a proxy for conditions where recombination is strongly suppressed by an electric field.

Taken together, the results obtained in this study demonstrate that TPB-coated WLS fiber systems can efficiently detect scintillation light in gaseous noble detectors. The LCE measured here should, however, be regarded as an upper limit for a realistic large-scale TPC, since it does not include additional losses expected with cathode, anode, and gate grids, as in NEXT-100, or attenuation along meter-scale fibers. Nevertheless, the measured efficiencies indicate that WLS fiber readout can provide sufficient light sensitivity for the S1 detection of low-energy calibration signals. In particular, experience from NEXT-White yielded approximately 10 p.e. for the 41.5 keV decay of $^{83m}$Kr, while electroluminescence photon statistics were not a limiting contribution to the detector energy resolution. These results therefore suggest that WLS-fiber systems remain a viable and scalable low-channel-count solution for future high-pressure noble-gas TPCs, even if the achievable LCE in a full detector is somewhat lower than the value reported here.
In addition, a barrel of WLS fibers providing a LCE at the percent level could, in principle, replace photomultiplier tubes as the primary scintillation readout in HPGXe-TPCs while maintaining an equivalent energy resolution. Such an approach would significantly reduce mechanical and engineering complexity, since WLS fibers and SiPMs are intrinsically compatible with high-pressure operation, whereas PMTs require dedicated pressure-resistant housings and optical interfaces~\cite{NEXT:2025yqw}, while also being less radiopure~\cite{NEXT:2019rum, Janicsko-Csathy:2010uif}. 
 
More generally, the combination of scalability, reduced channel count, and stable performance makes this approach a promising solution for future high-pressure TPCs and other rare-event experiments requiring efficient and uniform scintillation-light readout.

\section{Conclusions}\label{sec:conclusions}
In this work we have experimentally investigated the performance of wavelength-shifting (WLS) optical fibers coated with tetraphenyl butadiene (TPB) for the detection of scintillation light in gaseous argon and xenon at pressures up to 8.5 bar. An elongated high-pressure vessel instrumented with TPB-coated WLS fibers and SiPM readout was operated under well-controlled thermal and gas-purity conditions and characterized using both a $\mathrm{^{241}Am}$ source and tagged cosmic-ray muons. 

Using alpha-induced scintillation and correcting for geometrical shadowing and VUV absorption effects via dedicated Monte Carlo simulations, we measured a light-collection efficiency that is constant over the explored pressure range and equal to $1.07 \pm 0.01~\mathrm{(sta.)}~^{+0.06}_{-0.08}~(\mathrm{sys})~\%$ and $1.18 \pm 0.01~\mathrm{(sta.)}~^{+0.07}_{-0.09}~(\mathrm{sys})~\%$ for argon and xenon, respectively. These values are in good agreement with existing measurements in LAr~\cite{Anfimov:2020aev}. 

In addition, using an independent experimental setup, we performed a complementary measurement of the LCE employing PMT readout and wavelength-shifting fibers from a different manufacturer. The measurement, performed at 1~bar, yields $0.45 \pm 0.01~\mathrm{(sta.)} \pm 0.05~(\mathrm{sys.})~\%$. After accounting for the different photon detection efficiencies of the PMTs and SiPMs, this result is in good agreement with the corresponding SiPM-based measurement. This provides an important cross-check of the LCE determination using an independent readout configuration. However, since the two measurements rely on fibers with different optical properties and potentially different fiber-photosensor coupling conditions, the comparison should not be interpreted as a direct isolation of the photosensor technology or fiber manufacturer effects.

Complementary measurements with through-going cosmic muons in xenon enabled the determination of the scintillation light yield as a function of pressure in the absence of source-induced geometrical effects. These results demonstrate that, in the low-ionization regime where recombination is negligible, the scintillation light yield remains essentially independent of the operating pressure.

By combining the light yield measured at low pressure with the independently determined light-collection efficiency, we obtained a mean energy required to produce a scintillation photon of 
${w_{sc}^{\mathrm{Xe}} = 45\pm7~\mathrm{(sta.)}~^{+4}_{-5}~\mathrm{(sys.)}~\mathrm{eV}}$ at 1.5~bar, in agreement within the uncertainty with values reported in the literature under low-recombination conditions~\cite{Leardini:2021qnf,Henriques:2023evt}. This result, in addition to strengthening our LCE estimate, supports the view that the low-pressure cosmic-muon measurements are representative of HPGXe-TPC operation under a drift field, where electron recombination is negligible.

\begin{acknowledgements}
Fibers coating with TPB was performed at the University of Edinburgh thanks to the support of A.~Szelc. Front-end electronic boards are provided by I.~Sarra and S.~Ceravolo of the INFN LNF group. Cutting, polishing, and aluminization of WLS fibers was performed at LOMaC/LIP-Lisbon.

The NEXT Collaboration acknowledges support from the following agencies and institutions: the European Research Council (ERC) under Grant Agreements No. 951281-BOLD and 101039048-GanESS; the European Union’s Framework Programme for Research and Innovation Horizon 2020 (2014–2020) under Grant Agreement No. 860881-HIDDeN; the MCIN/AEI of Spain and ERDF A way of making Europe under grants PID2021-125475NB and RTI2018-095979, and the Severo Ochoa and Mar\'ia de Maeztu Program grants CEX2023-001292-S, CEX2023-001318-M and CEX2018-000867-S; the Generalitat Valenciana of Spain under grants PROMETEO/2021/087, ASFAE/2022/028, ASFAE/2022/029, CISEJI/2023/27, and CIDEXG/2023/16; the Department of Education of the Basque Government of Spain under the predoctoral training program non-doctoral research personnel; the Spanish la Caixa Foundation (ID 100010434) under fellowship code LCF/BQ/PI22/11910019; the Portuguese FCT under project UID/FIS/04559/2020 to fund the activities of LIBPhys-UC; the Israel Science Foundation (ISF) under grant 1223/21; the Pazy Foundation (Israel) under grants 310/22, 315/19 and 465; the US Department of Energy under contracts number DE-AC02-06CH11357 (Argonne National Laboratory), DE-AC02-07CH11359 (Fermi National Accelerator Laboratory), DE-FG02-13ER42020 (Texas A\&M), DE-SC0019054, DE-SC0019223 and DE-SC0024438 (Texas Arlington); the US National Science Foundation under award number NSF CHE 2004111; the Robert A Welch Foundation under award number Y-2031-20200401; the Dpt. of Science, Universities and Innovation of the Basque Government through the 2025 grant to the DIPC Neutrino Physics Laboratory. Finally, we are grateful to the Laboratorio Subterr\'aneo de Canfranc for hosting and supporting the NEXT experiment.

\end{acknowledgements}

\bibliographystyle{unsrturl}
\bibliography{biblio}

\end{document}

%% file: authors_epjc.tex
\author{S.R.~Soleti\orcid{0000-0002-5526-1414}\thanksref{a,addr2,addr17} 
\and
 S.~Torelli\orcid{0000-0003-3622-3524}\thanksref{b,addr2} 
 \and
 G.~Mart\'inez-Lema\orcid{0000-0002-1742-3531}\thanksref{addr4,addr7} 
\and
 H.~Almaz\'an\orcid{0000-0003-1772-2598}\thanksref{addr2} 
 \and
 A.~Beck\thanksref{addr25} 
\and
 A.~Castillo\orcid{0009-0004-1700-0979}\thanksref{addr2} 
 \and
 M.~del Barrio-Torregrosa\orcid{0009-0001-7036-6476}\thanksref{addr2,addr10} 
\and
 P.~Dietz\orcid{0009-0002-1803-0892}\thanksref{addr2} 
\and
 C.~Echeverria\thanksref{addr2} 
\and
 L.~Gurriana\thanksref{addr24}
% ...
}

\thankstext{a}{Corresponding author: \href{mailto:roberto.soleti@dipc.org}{roberto.soleti@dipc.org}}
\thankstext{b}{Corresponding author: \href{mailto:samuele.torelli@dipc.org}{samuele.torelli@dipc.org}}
\thankstext{f26}{Deceased.}
\thankstext{f35}{Now at Weizmann Institute of Science, Israel.}
\thankstext{f37}{On leave.}
\thankstext{f41}{NEXT Spokesperson.}
\thankstext{f89}{Now at University of North Carolina, USA.}
\thankstext{f90}{Now at Los Alamos National Laboratory, USA.}

\institute{
% ...
 Donostia International Physics Center, BERC Basque Excellence Research Centre, Manuel de Lardizabal 4, San Sebasti\'an / Donostia, E-20018, Spain \label{addr2} 
% ...
\and
 Unit of Nuclear Engineering, Faculty of Engineering Sciences, Ben-Gurion University of the Negev, P.O.B.~653, Beer-Sheva, 8410501, Israel \label{addr4} 
% ...
\and
 Instituto de F\'isica Corpuscular (IFIC), CSIC \& Universitat de Val\`encia, Calle Catedr\'atico Jos\'e Beltr\'an, 2, Paterna, E-46980, Spain \label{addr7} 
% ...
\and
 Department of Physics, Universidad del Pais Vasco (UPV/EHU), PO Box 644, Bilbao, E-48080, Spain \label{addr10} 
% ...
\and
 Ikerbasque (Basque Foundation for Science), Bilbao, E-48009, Spain \label{addr17} 
% ...
\and
 Laboratório de Instrumentação e Física Experimental de Partículas - LIP, Lisbon, 1649-003, Portugal \label{addr24}
 \and
 Nuclear Research Center Negev, Be’er-Sheva, Israel\label{addr25}
}